\pdfoutput=1
\documentclass[a4paper,12pt]{article}

\usepackage[a4paper,text={16.8cm,22.4cm}]{geometry}
\usepackage{amsmath,amssymb,bm,cite,graphicx}
\usepackage{booktabs}
\usepackage[normalem]{ulem}

\usepackage[Q=yes,pverb-linebreak=no]{examplep}

\allowdisplaybreaks 
\newcommand{\pTveto}{p_T^{\hspace{-0.3mm}\rm veto}\hspace{-0.3mm}}

\begin{document}

\begin{titlepage}

\begin{flushright}
CERN-PH-TH-2014-268\\
MITP/14-099\\
December 29, 2014%\\
%v2: April 2, 2015
\end{flushright}
 
\vspace{0.3cm}
\begin{center}
\Large\bf\boldmath
Automated NNLL+NLO Resummation\\ for Jet-Veto Cross Sections
\end{center}

\vspace{0.3cm}
\begin{center}
Thomas Becher$^a$, Rikkert Frederix$^b$, Matthias Neubert$^{c,d}$ and Lorena Rothen$^a$ \\
\vspace{0.4cm}
{\sl ${}^a$\,Albert Einstein Center for Fundamental Physics, Institut f\"ur Theoretische Physik\\
Universit\"at Bern, Sidlerstrasse 5, CH--3012 Bern, Switzerland\\[0.4cm]
${}^b$PH Department, TH Unit, CERN, CH-1211 Geneva 23, Switzerland\\[0.4cm]
${}^c$\,PRISMA Cluster of Excellence \& Mainz Institut for Theoretical Physics\\
Johannes Gutenberg University, D-55099 Mainz, Germany\\[0.4cm]
${}^d$\,Department of Physics, LEPP, Cornell University, Ithaca, NY 14853, U.S.A.}
\end{center}

\vspace{0.2cm}
\begin{abstract}
\noindent 
In electroweak-boson production processes with a jet veto, higher-order corrections are enhanced by logarithms of the veto scale over the invariant mass of the boson system. In this paper, we 
resum these Sudakov logarithms at next-to-next-to-leading logarithmic (NNLL) accuracy and match our predictions to next-to-leading order (NLO) fixed-order results. We perform the calculation in an automated way, for arbitrary electroweak final states and in the presence of kinematic cuts on the leptons produced in the decays of the electroweak bosons. The resummation is based on a factorization theorem for the cross sections into hard functions, which encode the virtual corrections to the boson production process, and beam functions, which describe the low-$p_T$ emissions collinear to the beams. The one-loop hard functions for arbitrary processes are calculated using the MadGraph5\Q{_}aMC@NLO framework, while the beam functions are process independent. We perform the resummation for a variety of processes, in particular for $W^+ W^-$ pair production followed by leptonic decays of the $W$ bosons. 
\end{abstract}
\vfil

\end{titlepage}

\section{Introduction}

In many experimental measurements a veto on hard jets is imposed to suppress backgrounds. Such a veto is particularly useful to suppress top-quark backgrounds to processes involving $W$ bosons, since the $W$ bosons from the decay of the top quarks come in association with $b$-jets, which are rejected by the jet veto. For example, a jet veto is crucial to measure Higgs production with subsequent decay $H\to W^+W^-$. It is imposed by rejecting events which involve jets with transverse momentum above a scale $\pTveto$, which is typically chosen to be $\pTveto\approx 20-30\,{\rm GeV}$. Since the veto scale is much lower than the invariant mass $Q$ of the electroweak final state, perturbative corrections to the cross section are enhanced by Sudakov logarithms of the ratio  $\pTveto/Q$. There has been a lot of theoretical progress over the past two years concerning the resummation of jet-veto logarithms in Higgs-boson production. Using the CAESAR formalism \cite{Banfi:2004yd}, these logarithms were first computed at next-to-leading logarithmic (NLL) order in \cite{Banfi:2012jm}, and this treatment was later extended to NNLL \cite{Banfi:2012yh}. In between these papers, an all-order factorization formula derived in Soft-Collinear Effective Theory (SCET) \cite{Bauer:2000yr,Bauer:2001yt,Beneke:2002ph,Becher:2014oda} was proposed \cite{Becher:2012qa}, and a resummed result which includes almost all of the ingredients required for N$^3$LL accuracy was presented \cite{Becher:2013xia}. A third group of authors performed an independent analysis in SCET \cite{Stewart:2013faa} and also combined the results for different jet multiplicities \cite{Liu:2012sz,Liu:2013hba,Boughezal:2013oha}. 

The jet veto is not only necessary in $H\to W^+ W^-$ but also in the measurement of the diboson cross section itself. The fact that LHC measurements \cite{ATLAS:2012mec,Chatrchyan:2013yaa,Chatrchyan:2013oev,ATLAS-CONF-2014-033} yield values of the $W^+ W^-$ cross section that are higher than theoretical predictions has triggered discussions as to whether this excess could be due to New Physics \cite{Curtin:2014zua,Kim:2014eva,Luo:2014fva}. To be sure whether there indeed is an excess, it is important to have reliable theoretical predictions not only for the total cross section, for which the next-to-next-to leading order (NNLO) result has been obtained recently \cite{Gehrmann:2014fva}, but also for the cross section in the presence of experimental cuts, most importantly in the presence of a jet veto.\footnote{Preliminary NNLO results for the rate in the presence of cuts were presented at a recent conference \cite{TalkMG}.} 
Several recent papers have addressed this issue and have come to somewhat different conclusions. In \cite{Jaiswal:2014yba}, the Sudakov logarithms associated with the jet veto were resummed at NNLL accuracy. It was claimed that resummation effects increase the cross section and bring the Standard-Model prediction in agreement with the experimental measurements. On the other hand, based on a study of transverse-momentum resummation, the authors of \cite{Meade:2014fca} concluded that resummation effects are small for the relevant values of $\pTveto$. Most recently, the effect of using a matched parton shower to predict the fiducial cross section, as it is done in the experimental analyses, was analyzed in \cite{Monni:2014zra}. These authors concluded that resummation effects are small and that a fixed-order computation of the fiducial rate would lead to theoretical predictions in agreement with the measurements, but that the matched parton shower overestimates the Sudakov suppression of the rate and leads to systematically lower theoretical predictions when extrapolating back to the total rate. 

In the present paper, we present an automated method to perform resummations for arbitrary vector-boson production processes involving jet vetoes. Instead of computing resummed cross sections analytically, on a case-by-case basis, we obtain them in an automated way using the MadGraph5\Q{_}aMC@NLO framework \cite{Alwall:2014hca}. Our method yields results which are accurate at NNLL and are matched to NLO fixed-order results. Such an automated procedure is obviously much more efficient and less error prone than computing the ingredients by hand or extracting them from the literature. Most importantly, our approach allows us to also include the decay of the vector bosons, along with cuts on the leptons in the final state. 

We have implemented two different methods to perform the resummation. The first one is based on reweighting tree-level events generated by MadGraph. It yields jet-veto cross sections accurate at NNLL order. The event weight includes universal resummation factors as well as the process-specific one-loop virtual corrections, which are computed using Madgraph5\Q{_}aMC@NLO. In the second method, we modify the NLO fixed-order computation in such a way that the end result is accurate at both NNLL and NLO. In this second method not only the hard function, which encodes the virtual corrections, but also the beam functions, which describe the emissions at small transverse momentum, are computed by Madgraph5\Q{_}aMC@NLO. 

Our paper is organized as follows. We start in Section~\ref{sec:fact} by reviewing the resummation formula for cross sections in the presence of a jet veto. We also discuss non-perturbative corrections and point out that they could be sizable, similar in magnitude as the recently calculated NNLO corrections. We then explain in Section~\ref{sec:reweight} how the automated resummation can be implemented in the Madgraph5\Q{_}aMC@NLO framework. In Section~\ref{sec:results} we use our method to compute cross sections for different boson-production processes and discuss in detail the scale and scheme choices and the resulting theoretical uncertainties. We compare our resummed predictions to fixed-order results for the cross sections, to the results obtained from a matched parton shower, and to the NNLL results of \cite{Jaiswal:2014yba}. We also match our resummed result to fixed-order NLO predictions. The relevant matching corrections turn out to be very small, which indicates that the bulk of the NLO result is already captured by the factorization formula evaluated with NNLL accuracy. This remains true after imposing cuts on the leptonic final state in the decays of the electroweak bosons. We compare predictions for the final states $Z$, $W^+W^-$ and $W^+W^-W^+$ and consider ratios of cross sections, which have small uncertainties if they are properly defined. We then discuss the implications of our results on the value of the $W^+W^-$ cross section and conclude in Section~\ref{sec:conclusion}.

\section{Factorization Theorem for Jet-Veto Cross Sections}
\label{sec:fact}

\begin{figure}[t!]
\centering
\includegraphics[width=0.65\textwidth]{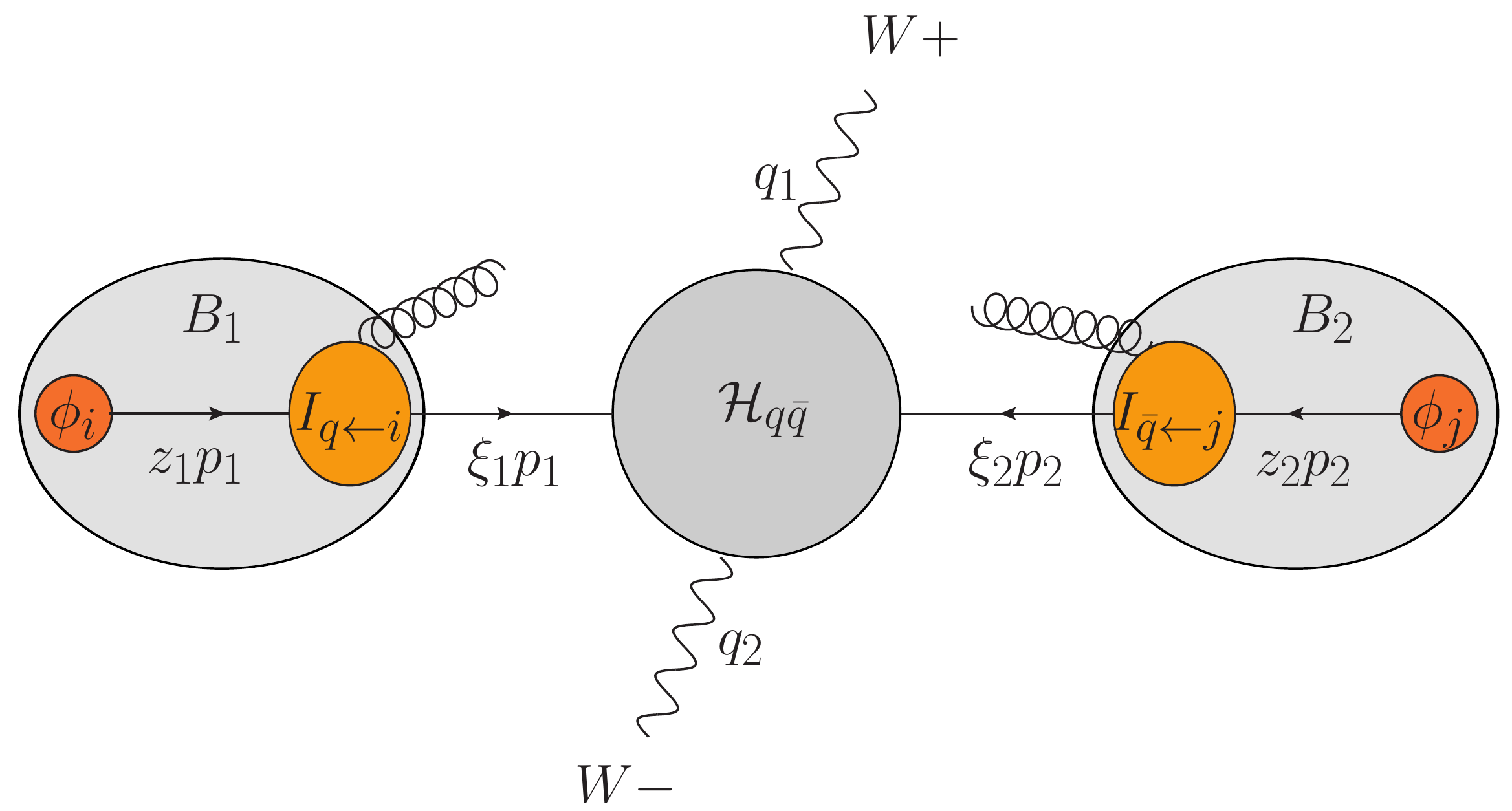}
\caption{Structure and kinematics of the factorization theorem for the $W^+ W^-$ production cross section in the presence of a jet veto.}
\label{fig:kin}
\end{figure}

We focus on electroweak-boson production processes with a veto on jets with transverse momentum above a cut $\pTveto$. The large logarithms which arise in the presence of the jet veto have the form $\alpha_s^n \ln^m(\pTveto/Q)$ with $m\leq 2n$, where $Q$ denotes the invariant mass of the boson system. Our goal is the resummation of these logarithms to all orders in perturbation theory and at leading power in the small ratio  $\pTveto/Q$. For concreteness, we will discuss the resummation for $W^+ W^-$ pair production in the following, but the formalism applies to any number of massive vector bosons and Higgs bosons or other massive color-singlet particles in the final state. The resummation is based on a factorization theorem which arises in the limit $\pTveto/Q \to 0$ \cite{Becher:2012qa}. Its schematic form is shown in Figure \ref{fig:kin}. The main ingredients of the theorem are hard functions ${\cal H}_{ij}$, which encode the virtual QCD corrections to the partonic hard-scattering processes $i+j\to W^+ W^-$, and two beam functions $\bar{B}_i$ and $\bar{B}_j$, which describe the low-$p_T$ emissions collinear to the two beams. 

Before writing out the factorization theorem in more detail, let us specify the kinematics of the process at low $\pTveto$. The momenta of the incoming protons are $p_1$ and $p_2$. The partons emerging from the parton distribution functions (PDFs) carry momenta $z_1 p_1$ and $z_2 p_2$. After possible emissions (described by the beam functions $\bar{B}_i$), the momenta $\xi_1 p_1$ and $\xi_2 p_2$ are left to produce the boson pair through a hard interaction $\mathcal{H}_{ij}$. In the limit of small transverse momenta we can neglect recoil effects, so that the partons are still collinear to the proton momentum after the emissions. We define
\begin{equation}
   \hat s = (q_1+q_2)^2=(\xi_1 p_1+\xi_2 p_2)^2=Q^2 \,, \qquad
   \hat t = (\xi_1 p_1-q_1)^2 \,, \qquad
   \hat u = (\xi_1 p_1-q_2)^2 \,,
\end{equation}
with $\hat s+\hat t+\hat u=2M_W^2$. Note that our definition of the variable $\hat s $ differs from the standard choice $(z_1 p_1+ z_2 p_2)^2$. The quantity $\hat{s}$ we define is the one relevant for the boson production process, i.e.\ the one that enters the hard function. In the small transverse-momentum limit of the emissions, we obtain
\begin{equation}\label{eq:kin}
\begin{aligned}
   \xi_1 &=\frac{\bar n\cdot q}{\bar n\cdot p_1}=\frac{Q}{\sqrt{s}}\,e^{-y}\qquad\Rightarrow\qquad   
    \xi_1p_1=\left(\bar n\cdot q\right)\frac{n}{2}\\ 
   \xi_2 &=\frac{n\cdot q}{n\cdot p_2}=\frac{Q}{\sqrt{s}}\,e^{y}\,\,\,\qquad\Rightarrow\qquad
    \xi_2p_2=\left(n\cdot q\right)\frac{\bar n}{2}\,,
\end{aligned}
\end{equation}
where $n^\mu=(1,0,0,1)$ and $\bar{n}^\mu=(1,0,0,-1)$ are two light-cone vectors in the beam directions, $y$ denotes the rapidity of $q=q_1+q_2$ in the laboratory frame, and $s=(p_1+p_2)^2$. The crucial feature of \eqref{eq:kin} is that it shows that one can obtain the arguments of the hard function directly from the vector-boson (and proton) kinematics. The same is true for an arbitrary electroweak final state.

At low $\pTveto$, the differential cross section in the presence of a jet veto has the factorized form \cite{Becher:2012qa,Becher:2013xia} 
\begin{equation}\label{eq:sigma}
   \frac{d^3\sigma(\pTveto)}{dy\,dQ^2\,d\hat t}
   =\sum_{i,j=g,q,\bar{q}} \sigma^0_{ij}(Q^2,\hat t,\mu)\,P_{ij}(Q^2,\hat t,\pTveto,\mu)\,
    \bar B_i(\xi_1,\pTveto)\,\bar B_j(\xi_2,\pTveto) \,.
\end{equation}
Here $i$ and $j$ are the flavors of the partons which enter the hard-scattering process after initial-state radiation, and $\sigma^0_{ij}(Q^2,\hat t)$ is the Born-level cross section for the production of the electroweak final state. Since the electroweak final state is a color singlet, we either deal with $q\bar{q}$ or $gg$. For $W^+ W^-$ pair production at leading order only the quark channels contribute, but starting from NNLO also the gluon-induced reaction occurs. 

The second ingredient in \eqref{eq:sigma} are the beam functions $\bar B_i(\xi,\pTveto)$, which are given by a convolution of a perturbative kernel $\bar{I}_{q\leftarrow k}(z,\pTveto,\mu)$ describing the emissions with the standard PDFs $\phi_k$: 
\begin{equation}\label{pdfmatch}
   \bar B_i(\xi,\pTveto) = \sum_{k=g,q,\bar q} \int_\xi^1\!\frac{dz}{z}\,
    \bar I_{i\leftarrow k}(z,\pTveto,\mu)\,\phi_k(\xi/z,\mu) \,.
\end{equation}
The bar over these functions indicates that a factor $e^{h_i(\pTveto,\mu)}$ has been extracted from the original definitions of these functions in terms of SCET operators, called $B_i$ and $I_{i\leftarrow k}$, such that 
\begin{equation}\label{eq:convol}
   \bar B_i(\xi,\pTveto) = e^{-h_i(\pTveto,\mu)} B_i(\xi,\pTveto,\mu) \,,
\end{equation}
and analogously for $\bar{I}_{i\leftarrow k}(z,\pTveto,\mu)$. This factor is normalized such that $h_i(\pTveto,\pTveto)=1$, and chosen such that the remaining function $\bar B_i(\xi,\pTveto)$ is renormalization-group (RG) invariant. The explicit form of $h_i(\pTveto,\mu)$ as well as the one-loop kernels $\bar I_{i\leftarrow k}(z,\pTveto,\mu) $ are listed in the appendix.

The final ingredient in \eqref{eq:sigma} is the prefactor $P_{ij}(Q^2,\hat t,\pTveto,\mu)$, which includes the hard function and the resummation of large logarithms. It has the form
\begin{equation}\label{eq:fact}
   P_{ij}(Q^2,\hat t,\pTveto,\mu)
   = \mathcal{H}_{ij}(Q^2,\hat t,\mu_h)\, E_{i}(Q^2,\pTveto,\mu_h,\mu, R) \,,
\end{equation}
where the hard function 
\begin{equation}
    \mathcal{H}_{ij}(Q^2,\hat t,\mu_h) 
    = 1+\frac{\alpha_s(\mu_h)}{4\pi}\,\mathcal{H}_{ij}^{(1)}(Q^2,\hat t,\mu_h) + \dots
\end{equation}
contains higher-order finite virtual corrections to the Born-level cross section. Since these higher-order corrections contain (double) logarithms of $Q/\mu_h$, the hard matching scale $\mu_h$ should be chosen of order $Q$. The evolution of the hard function to a lower scale $\mu\ll Q$ is controlled by an RG evolution equation. The corresponding evolution function $U_{i}(Q^2,\mu_h,\mu )$, together with the collinear anomaly \cite{Becher:2010tm} and the prefactors extracted from the beam functions, is absorbed into the factor $E_i$ in \eqref{eq:fact}. The collinear anomaly arises due to light-cone divergences and provides an additional source of large logarithms in processes sensitive to small transverse momenta.
The explicit form of the quantity $E_i$ reads
\begin{equation}\label{eq:Prefactor}
   E_{i}(Q^2,\pTveto,\mu_h,\mu, R) 
   = U_{i}(Q^2,\mu_h,\mu ) \left( \frac{Q}{\pTveto} \right)^{-2F_{i}(\pTveto,\mu, R)} 
    e^{2h_{i}(\pTveto,\mu)} \,.
\end{equation}
The evolution factor at NNLL accuracy is given in the appendix. It differs for quark-initiated ($i=q$) and gluon-initiated ($i=g$) processes but is independent of the quark flavors. Note that the evolution factor depends on the kinematics of the final state only via the invariant mass $Q$. The anomaly exponent $F_{i}(\pTveto,\mu, R)$ resums the large anomalous logarithms in the beam and soft functions, which arise from the rapidity difference between the modes which contribute to the individual functions \cite{Becher:2010tm,Chiu:2011qc,Chiu:2012ir}. Starting from two-loop order (which is needed for NNLL resummation) this exponent depends on the jet radius $R$, but it is the same for any $k_T$-style sequential jet-clustering algorithm. The explicit form of the two-loop exponent can be found in the appendix. It was calculated in \cite{Becher:2013xia} and is related to the function $\cal F$ obtained earlier in \cite{Banfi:2012yh}.

We stress that the factorization theorem holds up to power corrections suppressed by $\pTveto/Q$, and up to nonperturbative effects suppressed by $\Lambda_{\rm QCD}/\pTveto$. For the weak-boson transverse-momentum spectrum, these corrections depend on $\vec{p}_T^2$ and hence are of second order in $\pTveto/Q$ and $\Lambda_{\rm QCD}/\pTveto$. The definition of the jet veto, on the other hand, involves an absolute value of the jet transverse momentum, and for this reason there can be first-order power corrections. Non-perturbative corrections to processes involving an anomaly were studied in \cite{Becher:2013iya}, where it was found that these effects are enhanced by a logarithm of the rapidity difference between the left- and right-collinear emissions and can be viewed as a non-perturbative contribution to the anomaly exponent $F_i$ in (\ref{eq:Prefactor}). The leading non-perturbative corrections to jet-veto cross sections are therefore expected to scale as 
\begin{equation}\label{eq:NP}
   \sigma_{\rm NP}(\pTveto)\sim \sigma^0\times\frac{\Lambda_{\rm NP}}{\pTveto}\,
    \ln\frac{Q}{\pTveto} \,.
\end{equation}
Due to the fact that the correction is of first order and logarithmically enhanced, these effects might not be negligible. For example, assuming $\Lambda_{\rm NP} = 0.5\,{\rm  GeV}$ and $\pTveto = 20 \,{\rm  GeV}$, one ends up with a 6\% effect at  $Q=222\,{\rm GeV}$, which is the median $Q$ value in $W^+ W^-$ production. Numerically, this is not much smaller than the NNLO correction to the total cross section calculated in \cite{Gehrmann:2014fva}. The value of the non-perturbative quantity $\Lambda_{\rm NP}$ is unknown, but it could be obtained from the matrix element 
\begin{equation}\label{eq:pTmat}
   \mathcal{M}_{\rm veto}
   = \sum\hspace{-0.65cm}\int\limits_{X,{\rm reg}} p_T^{\rm jet}
    \left| \langle X | S_n^\dagger(0)\,S_{\bar{n}}(0) |0\rangle \right|^2
\end{equation}
of two soft Wilson lines along the beam directions, where $p_T^{\rm jet}$ is the transverse momentum of the leading jet in the final state $X$. The phase-space integrals in the matrix element $\mathcal{M}_{\rm veto}$ suffer from a rapidity divergence, which needs to be regularized. The parameter $\Lambda_{\rm NP}$ multiplies the rapidity divergence (see \cite{Becher:2013iya} for more details). To get an idea of the size of non-perturbative effects, we have computed the hadronization effects to the cross section using Pythia 8 \cite{Sjostrand:2014zea} with its default tune. We find that they change the cross section by about 10\% at $\pTveto=10\,{\rm GeV}$ and 3\% at $\pTveto=20\,{\rm GeV}$. Above $\pTveto>20\,{\rm GeV}$, the simple parametrization in \eqref{eq:NP} with $\Lambda_{\rm NP}=240\,{\rm MeV}$ provides a good description of the Pythia hadronization corrections, while a first-order power correction without logarithmic enhancement would underestimate the effects at higher $\pTveto$ values. However, one should be careful in relying on Pythia hadronization effects in the context of precision calculations. There are other examples, such as the event-shape variable thrust, where Pythia appears to underestimate the size of these effects \cite{Abbate:2010xh}. In the absence of a non-perturbative evaluation of the soft matrix element \eqref{eq:pTmat}, the only reliable way to determine the size of the power corrections is to measure jet-veto cross sections at several different low $\pTveto$ values and for different values of $Q$ and compare it to the resummed perturbative prediction. It would be interesting to do so, and there should be enough Drell-Yan  and $Z$-production data to make such a study possible.

\section{Automated Resummation}
\label{sec:reweight}

We now explain how to automate the resummation by suitably modifying existing fixed-order results. We shall employ two different resummation schemes. In {\it Scheme A}, we work with tree-level events obtained from MadGraph5\Q{_}aMC@NLO \cite{Alwall:2014hca}. We supply the beam functions from explicit calculations but compute the hard functions automatically and then reweight the events to achieve the resummation. In {\it Scheme B}, we use MadGraph5\Q{_}aMC@NLO in fixed-order mode and compute the NLO cross section with a jet veto. To achieve the resummation we subtract the logarithmically enhanced pieces from the fixed-order cross section and multiply them back in resummed form. In this second scheme, both the hard functions and the beam functions are computed using  MadGraph5\Q{_}aMC@NLO. The second scheme is more convenient for practical computations but limited to NNLL order, while the first scheme allows (in principle) for arbitrary accuracy of the resummation.

\subsection{Scheme A: NNLL from Reweighting Born-Level Events\label{sec:schemeA}}

The fact that the resummed result \eqref{eq:sigma} has Born-level kinematics in the limit $\pTveto\to 0$ makes it possible to achieve the resummation of large logarithms by a simple reweighting procedure. If we use a tree-level event generator such as MadGraph, the resummation can be implemented by rescaling the event weights with the ratio of the resummed to the tree-level cross sections at each kinematic point. Specifically, we need to replace the PDFs $\phi_i$ used in the leading-order (LO) result with the beam functions $\bar B_i$, and we need to supply the hard matching correction and the resummation factor $E_i$. For an incoming particle of flavor $i,j \in \{q, \bar{q}, g\}$, the reweighting factor at NNLL order reads
\begin{equation}\label{NNLLweight}
\begin{aligned}
   d\sigma_{ij}^{\text{NNLL}}(\pTveto) 
   &=\left(1+\frac{\alpha_s(\mu_h)}{4\pi}\mathcal{H}_{ij}^{(1)}(Q^2,\hat t,\mu_h)\right)\,
    E_{i}(Q^2,\pTveto,\mu_h,\mu, R)\\
   &\quad\times\frac{\bar{B}_i(\xi_1,\pTveto,\mu)}{\phi_i(\xi_1,\mu_{\text{Mad}})}\,
    \frac{\bar{B}_{j}(\xi_2,\pTveto,\mu)}{\phi_j(\xi_2,\mu_{\text{Mad}})}
    \left(\frac{\alpha_s(\mu)}{\alpha_s(\mu_{\text{Mad}})}\right)^N 
    d\sigma^{0}_{ij}(\mu_{\text{Mad}}) \,.
\end{aligned}
\end{equation}
All the kinematic variables are determined by the event kinematics. At leading order $\xi_1$ and $\xi_2$ are just the momentum fractions of the incoming particles, $\xi_i=2\,E_i/\sqrt{s}$. Note that we do not need to adopt the same value of the renormalization scale $\mu$ as in the Born-level events, which were evaluated at a scale $\mu_{\text{Mad}}$ inherent to the MadGraph code. However, in cases such as Higgs production, where the Born-level cross section depends on $\alpha_s$, we have to multiply by the appropriate power $N$ of the ratio $\alpha_s(\mu)/\alpha_s(\mu_{\text{Mad}})$, where $N=2$ for gluon-induced processes. We therefore only run MadGraph once, with a fixed reference scale $\mu_{\text{Mad}}$. Scale uncertainties can then be estimated by repeating the reweighting with different values of $\mu$ and $\mu_h$.

Let us now detail the numerical implementation of the reweighting factor, starting with the beam functions, which are defined in (\ref{pdfmatch}) in terms of convolutions of perturbative kernel functions with PDFs. At one-loop order, they are linear in the logarithm of $\pTveto$, and hence
\begin{equation}
   \bar{B}_i(\xi,\pTveto,\mu) = \phi_i(\xi,\mu) + \frac{\alpha_s(\mu)}{4\pi} 
    \left( b_i(\xi,\mu) + c_i(\xi,\mu) \ln\frac{\mu}{\pTveto} \right) .
\end{equation}
To perform the reweighting in an efficient way, we compute and tabulate the convolution integrals for $b_i(\xi,\mu)$ and $c_i(\xi,\mu)$ for a grid of $\xi$ and $\mu$ values. Since the beam functions are independent of the final state, this can be done once and for all. Using the same grid as the underlying PDFs itself, we then use standard PDF interpolation routines to have fast and accurate numerical representations for the beam functions. We have implemented the beam functions and the resummation factor $E_i(Q^2,\pTveto,\mu_h,\mu, R)$ in a small Fortran code, which is called by the event reweighting routine written in Python. 

The most complicated component of the reweighting factor by far is the hard function $\mathcal{H}_{ij}^{(1)}(Q^2,\hat t,\mu_h)$. This is process dependent and its computation requires a one-loop calculation. Fortunately, the necessary one-loop computations have been automated in the past few years. In particular, the MadGraph5\Q{_}aMC@NLO framework provides the possibility to evaluate virtual corrections at specific phase-space points \cite{Hirschi:2011pa}. We use this code to evaluate the virtual corrections $V_{ij}$ for each event. At each phase-space point, the code provides the result in the form of the coefficients $C_i$ of the double pole, single pole, and finite terms in the expansion in $\epsilon$, which is written in the form
\begin{equation}
   V_{ij} = d\sigma_{ij}^0(\mu) \left[ 1 + \frac{\alpha_s(\mu)}{4\pi} 
    \frac{2\,e^{-\epsilon \gamma_E}}{\Gamma(1+\epsilon)} 
    \left(\frac{\mu^2}{\mu_{\rm Mad}^2}\right)^\epsilon 
    \left(\frac{C_2}{\epsilon^2} + \frac{C_1(\mu_{\rm Mad})}{\epsilon} + C_0(\mu_{\rm Mad}) 
    \right)_{\!ij} \right] .
\end{equation} 
The scale $\mu_{\rm Mad}$ can be chosen when running the MadLoop code. In the $q\bar{q}$ channel, the double-pole coefficient is $C_2=-C_F\gamma^{\rm cusp}_0/2=-2C_F$, while the other two coefficients depend on the choice of $\mu_{\rm Mad}$. For $\mu_{\rm Mad}=Q$, the coefficient of the single-pole term is $C_1(Q)=\gamma^q_0=-3C_F$, and the finite part in the expansion of the above expression in $\epsilon$ directly yields the hard function
\begin{equation}
   \mathcal{H}_{q\bar{q}}^{(1)}(Q^2,\hat t,\mu)  
   = 2 C_0(Q) + C_F \left(\frac{\pi ^2}{3}-2 \ln^2\frac{Q^2}{\mu^2}+ 6 \ln\frac{Q^2}{\mu^2} \right) .
\end{equation}
For $Z$-boson production one has $C_0(Q)=-32/3+4\pi^2/3$. For other choices $\mu_{\rm Mad}\ne Q$ this result gets modified to
\begin{equation}
   \mathcal{H}_{q\bar{q}}^{(1)}(Q^2,\hat t,\mu) 
   = 2 C_0(\mu_{\rm Mad}) + C_F \left[ \frac{\pi^2}{3}+2\ln^2\frac{\mu_{\rm Mad}^2}{\mu^2}
    + \ln\frac{\mu_{\rm Mad}^2}{\mu^2} \left(6-4\ln\frac{Q^2}{\mu^2}\right) \right] .
\end{equation}
In practice, we first compute the hard function at some value of the reference scale $\mu_{\rm Mad}$ for each event and write the result in the event record. The result at a different scale can then be obtained using the above relation. The reweighting script uses the result for the hard function and combines it with the beam functions and the resummation factor.

To obtain the best possible prediction, we match our result to the NLO fixed-order result for the cross-section. This matching allows us to also include terms which are power suppressed as $\pTveto \to 0$. The simplest way to achieve the matching is to subtract from the resummed result its expansion to NLO and to then add back the full NLO result
\begin{equation}\label{eq:match}
   \frac{d\sigma^{\rm NNLL+NLO}}{d\pTveto} 
   = \frac{d\sigma^{\rm NNLL}}{d\pTveto} 
    - \left. \frac{d\sigma^{\rm NNLL}}{d\pTveto}\right|_{\text{expanded to NLO}} 
    + \frac{d\sigma^{\rm NLO}}{d\pTveto} \,.
\end{equation}
Our final NNLL+NLO result resums higher-order terms that are logarithmically enhanced, but also includes the full NLO result. To obtain the expansion of the resummed result, we simply do the reweighting with the fixed-order expansion of the reweighting factor in \eqref{NNLLweight}. The NLO result can be obtained from running MadGraph5\Q{_}aMC@NLO in fixed-order mode. The difference between the full NLO result and the expansion of the resummed result is called the matching correction. By definition, this correction vanishes as $\pTveto\to 0$ and is expected to scale as $\pTveto/Q$. As we will discuss in Section~\ref{sec:matching}, it is numerically very small for the values of $\pTveto$ which are experimentally relevant.

\subsection{Scheme B: NNLL+NLO with Automated Computation of the Beam Functions and Matching Corrections}

In the reweighting scheme discussed above, we use MadGraph5\Q{_}aMC@NLO to compute the hard functions but supply the beam functions from an explicit calculation. One can go even further and also compute the beam functions and the matching corrections automatically and in a single step. This is done by first factoring out the hard corrections and then performing a NLO run in the presence of the jet veto. An advantage of this second approach is that the beam functions are computed on the fly and it is therefore easy to use different PDF sets without any need to recompute the beam functions. A slight disadvantage is that one has to run MadGraph5\Q{_}aMC@NLO in NLO mode. One can thus no longer work with events and will have to perform a new run when changing the cuts. However, if the matching is included in Scheme A described above, then a NLO run is needed also in this case. Note also that Scheme B only works at NNLL accuracy, while Scheme A allows for arbitrary precision if the necessary reweighting factor is supplied.

In order not to contaminate the matching corrections with the large logarithms contained in the hard function, we factor out the prefactor $P_{ij}$ in \eqref{eq:fact} and define a reduced cross section $\tilde\sigma_{ij}$ by
\begin{equation}
   d\sigma_{ij}(\pTveto) = P_{ij}(Q^2,\hat t,\pTveto)\,d\tilde\sigma_{ij}(\pTveto) \,.
\end{equation}
The reduced cross section has the form 
\begin{equation}
   d\tilde\sigma_{ij}(\pTveto) = d\sigma_{ij}^0(Q^2,\hat t,\mu)\,
    \bar B_i(\xi_1,\pTveto)\,\bar B_j(\xi_2,\pTveto) + \Delta\tilde\sigma \,,
\end{equation}
where $\Delta\tilde\sigma={\cal O}(\pTveto/Q)$ contains the power corrections and is given by the matching correction \eqref{eq:match} divided by the prefactor. The function $P_{ij}$ receives one-loop corrections from the hard function and the evolution factor $E_{i}$ so that we can write
\begin{equation}\label{eq:reduced}
\begin{aligned}
   d\tilde\sigma_{ij}(\pTveto) = d\sigma_{ij}^{\text{NLO}}(\pTveto,\mu)
    - \frac{\alpha_s(\mu)}{4\pi} \left( \mathcal{H}_{ij}^{(1)}(Q^2,\hat t,\mu)
    + E_{i}^{(1)}(Q^2,\pTveto,\mu) \right) d\sigma_{ij}^0(\mu) \,.
\end{aligned}
\end{equation}
Provided we choose $\mu\sim\pTveto$ in the reduced cross section $\tilde\sigma$, all large logarithms are resummed in the RG-invariant prefactor $P_{ij}$. Multiplying back the prefactor then yields the full NNLL+NLO cross section in the form
\begin{equation}\label{eq:reweight}
\begin{aligned}
   d\sigma_{ij}^{\text{NNLL+NLO}}(\pTveto)
   &= P_{ij}(Q^2,\hat t,\pTveto)\times d\tilde\sigma_{ij}(\pTveto)\\[0.5em]
   &= \left( 1+\frac{\alpha_s(\mu_h)}{4\pi}\mathcal{H}_{ij}^{(1)}(Q^2,\hat t,\mu_h) \right) 
    E_{i}(Q^2,\pTveto,\mu_h,\mu, R) \\
   &\quad\times \bigg[ d\sigma_{ij}^{\text{NLO}}(\pTveto,\mu) 
    - \frac{\alpha_s(\mu)}{4\pi} \left(\mathcal{H}_{ij}^{(1)}(Q^2,\hat t,\mu)
    + E_{i}^{(1)}(Q^2,\pTveto,\mu)\right) d\sigma_{ij}^{0}(\mu) \bigg] \,. \\[0.5em]
\end{aligned}
\end{equation}
Note that the matching procedure differs from the other scheme. In \eqref{eq:match} above, we performed a purely additive matching, while in \eqref{eq:reweight} the resummation factor $E_i$ appears as an overall factor. This multiplicative matching generates higher-order logarithmic terms also for the power-suppressed contributions of order $\pTveto/Q$ and higher. These additional terms are not controlled by the factorization theorem (\ref{eq:sigma}), which holds only at leading power, but one can hope that at least some of the logarithmic terms at subleading power are universal and will be captured by this treatment. For the case of Higgs production, the multiplicative matching scheme is preferred, since the perturbative corrections to the hard function are very large. In (\ref{eq:reweight}) they are extracted as a overall factor. For the $q\bar{q}$-initiated processes we study in this paper, the two schemes give almost indistinguishable results, as we will see in Section \ref{sec:matching} below.

To implement \eqref{eq:reweight} in MadGraph5\Q{_}aMC@NLO we have directly modified its Fortran code by including the logarithmically enhanced terms. The expanded logarithmically enhanced terms, i.e.~the second term on the right-hand side of \eqref{eq:reduced}, is similar to the compensating Sudakov factor introduced in the FxFx merging prescription, see (2.46) of \cite{Frederix:2012ps}, and it is therefore implemented at the same place in the code. In MadGraph5\Q{_}aMC@NLO each real-emission phase-space configuration has corresponding Born kinematics defined by the FKS mapping \cite{Frixione:1995ms}. Therefore we can always compute the prefactor $P_{ij}$ using Born kinematics, and it can multiply the complete reduced cross section, including the real-emission contributions. In order to improve the run time, the time-consuming one-loop matrix elements are computed only once for each phase-space configuration, cached in memory, and used also for the (expanded) hard function. However, compared to normal running of MadGraph5\Q{_}aMC@NLO, we cannot reduce the number of calls to the virtual corrections by using suitable approximations of it, as described in Sec.~2.4.3 of \cite{Alwall:2014hca}, because the reduced cross section is multiplied by them, resulting in positive feedback loops in setting up the approximations. When running MadGraph5\Q{_}aMC@NLO in fNLO mode, setting the parameter~\texttt{ickkw}~in the~\texttt{run\_card.dat}~to \texttt{-1} turns on in the inclusion of the logarithmically enhanced terms and sets the hard and soft scales to $Q$ and $\pTveto$ (given by the \texttt{ptj} parameter in the \texttt{run\_card.dat}), respectively. Hard and soft scale variations, as well as PDF uncertainties, can be computed at minimal CPU costs by reweighting~\cite{Frederix:2011ss}. This addition to the MadGraph5\Q{_}aMC@NLO will become public with the release of version 2.3 of the code. This version will also include the necessary scripts to perform the resummation using Scheme A described in Section \ref{sec:schemeA}.

\section{Phenomenological Results}
\label{sec:results}

We now proceed to give numerical results for different electroweak-boson production cross sections. Before presenting our final results, we discuss a variety of issues such as the proper choice of matching and factorization scales, the size of the matching corrections and the difference between the two resummation schemes discussed in the previous section. We then present results for the $W^+W^-$ cross section as well as the cross section including the decay of the $W$ bosons with cuts on the final-state leptons. Since the published measurements \cite{ATLAS:2012mec,Chatrchyan:2013yaa} were taken at $\sqrt{s} = 7\,{\rm TeV}$, we will present our results for this center-of-mass energy. For the electroweak parameters we use MadGraph5 default values, in particular $\alpha_{\rm em}=1/132.5$, $G_F=1.166\times 10^{-5}\,{\rm GeV}^{-2}$, $M_W=80.42\,{\rm GeV}$ and $M_Z=91.19\,{\rm GeV}$.

In all of our results below, we work with the MSTW2008NNLO PDF set and its associated value $\alpha_s(M_Z)=0.1171$ \cite{Martin:2009iq}. The choice of a NNLO PDF set seems appropriate, because we believe that the resummation captures the most important part of the NNLO corrections. In order to illustrate the size of the higher-order terms captured by resummation, we will also evaluate the NLO corrections using the NNLO PDF set, which increases the NLO prediction at $\pTveto = 20\, {\rm GeV}$ by about 2\% in the case of $W^+W^-$ production. We will define our jets using the anti-$k_T$ algorithm with a jet radius of $R=0.4$. The only quantity sensitive to the jet radius at NNLL+NLO accuracy is the anomaly exponent $F_{ij}(\pTveto,\mu,R)$, and it is the same for all $k_T$-style clustering algorithms. As the default scheme for our plots we use Scheme~A, since it is easier to disentangle and discuss the individual ingredients of the calculation (NLL versus NNLL resummation, matching to fixed-order perturbation theory) in this scheme. However, we find that both schemes give almost indistinguishable numerical results at NNLL+NLO level. 

\subsection{Resummed Results and Choice of the Hard Scale}

\begin{figure}[t!] 
\begin{tabular}{rr}
   \mbox{\hspace{4mm}\includegraphics[width=0.43\textwidth]{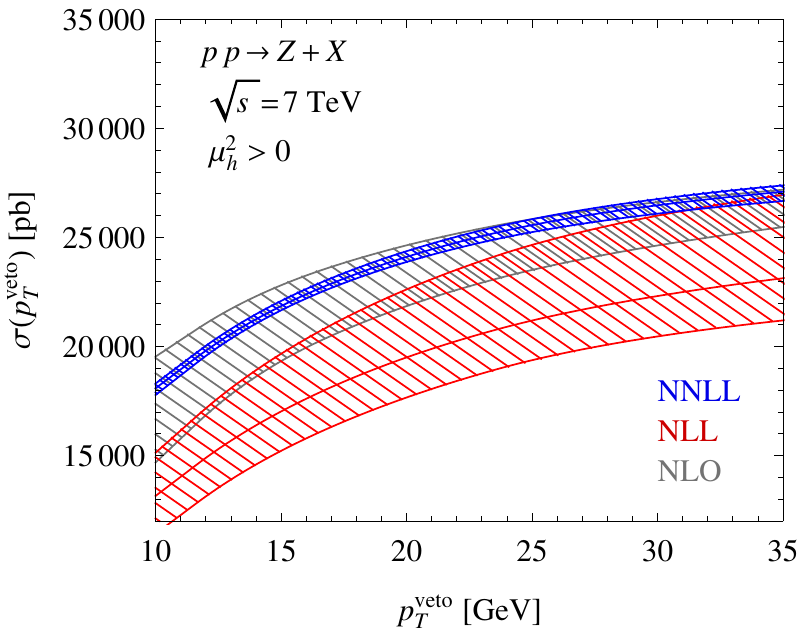}}&   
   \mbox{\includegraphics[width=0.43\textwidth]{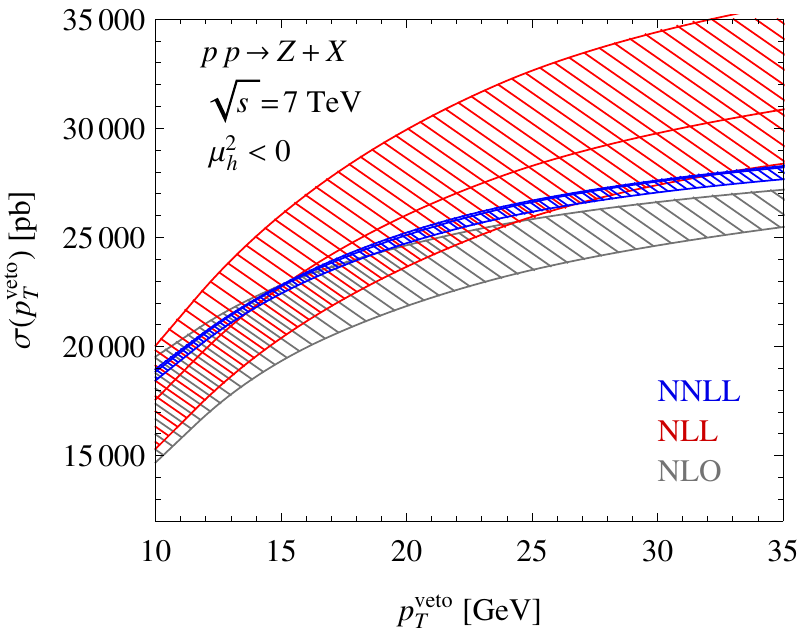}}\\
   \mbox{\includegraphics[width=0.4\textwidth]{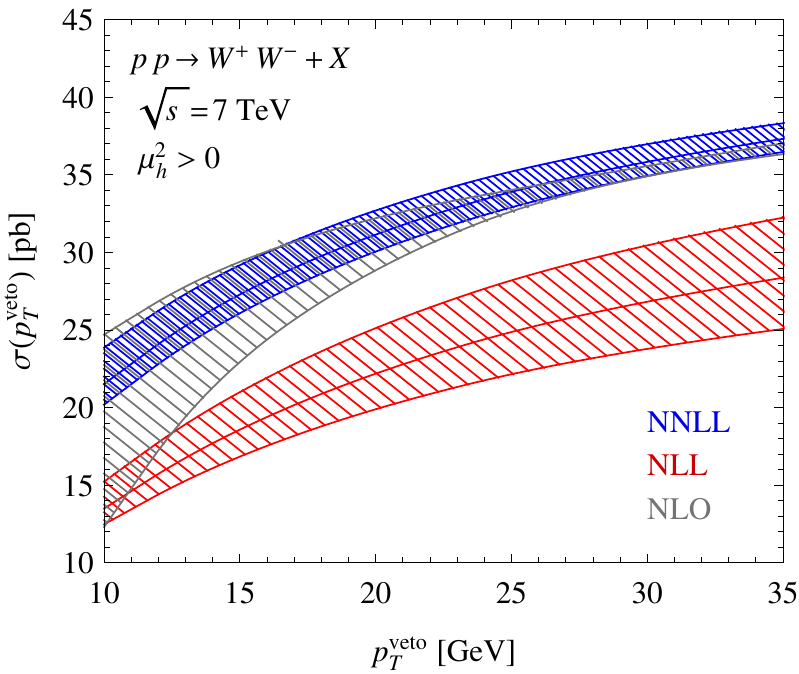}}& 
   \mbox{\includegraphics[width=0.4\textwidth]{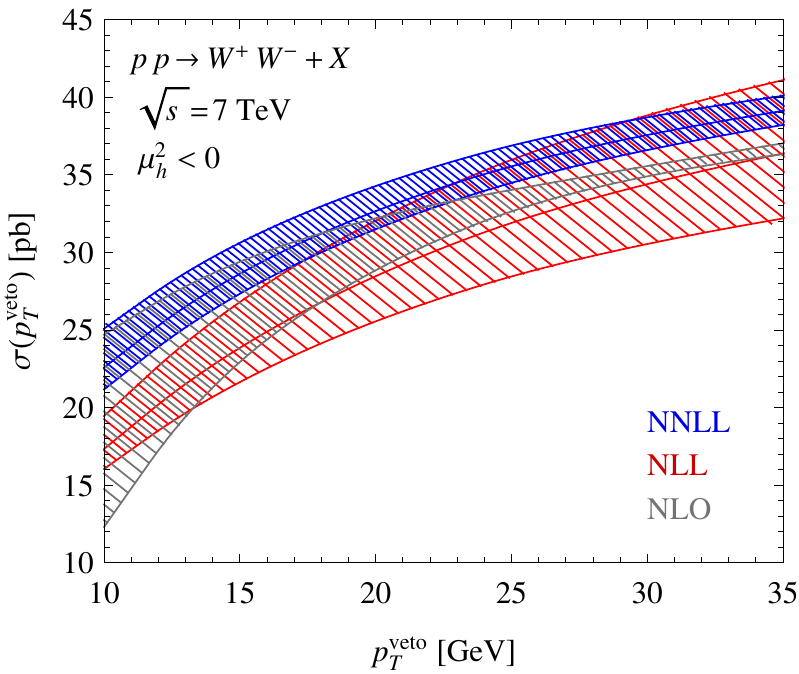}}\\
\end{tabular}
\caption{Resummed cross sections for $Z$-boson production (top) and $W^+ W^-$ pair production (bottom) obtained at NLL (red) and NNLL (blue) order. The bands are obtained by varying the hard matching scale $\mu_h$ and the factorization scale $\mu$ by factors of~2 about their default values $|\mu_h|=Q$ and $\mu=\pTveto$. The gray bands show the fixed-order NLO results with scale variation $\mu_r=\mu_f\in[\pTveto/2,\,2Q]$ for comparison. The panels on the left refer to the standard choice $\mu_h^2>0$, while those on the right show results obtained using $\mu_h^2<0$.}
\label{WWZRes}
\end{figure}

In Figure~\ref{WWZRes} we show the results for the resummed $Z$-boson and $W^+ W^-$ pair production cross sections at $\sqrt{s}=7$\,TeV, obtained with $n_f=5$ light quark flavors and jet radius parameter $R=0.4$. Here and below the two scales $\mu$ and $\mu_h$ are varied independently by factors of~2 about their default values $\mu=\pTveto$ and $\mu_h=Q$, where $Q$ is the invariant mass of the electroweak final state, i.e.\ $Q^2=M_Z^2$ for $Z$-boson production and $Q^2=(q_{1}+q_{2})^2$ for the $W^+ W^-$ final state (defined on an event-by-event basis). The resulting uncertainties are then added quadratically. In addition to the standard scale choice $\mu_h^2\approx Q^2$ we consider using an imaginary value for the hard matching scale, such that $\mu_h^2\approx -Q^2$. The corresponding results are shown on the right-hand side of Figure~\ref{WWZRes}. For comparison, we also show the NLO fixed-order results, which will be discussed in more detail in the next section. In all cases, we observe that going from NLL to NNLL accuracy improves the stability of the predictions significantly. Also, the NNLL bands are closer to the fixed-order NLO results than the NLL bands. 

The use of an imaginary value of the hard matching scale $\mu_h$ has been advocated in the context of Higgs production, because it maps the relevant hard function onto the space-like gluon form factor \cite{Ahrens:2008qu,Ahrens:2008nc}. This Euclidean quantity shows a much better perturbative behavior than the time-like form factor, which suffers from large numerical corrections $\sim(\alpha_s\pi^2)^n$ due to imaginary parts from Sudakov double logarithms, which arise in time-like kinematics. The same arguments apply to the case of $Z$ production. In \cite{Jaiswal:2014yba}, the choice $\mu_h^2<0$ was applied to $W^+ W^-$ pair production, and it was argued that this leads to a significant enhancement of the cross section, bringing the theoretical prediction in agreement with LHC measurements. Indeed, one can observe from Figure~\ref{WWZRes} that the resummed results for the cross sections obtained with $\mu_h^2<0$ are significantly larger than those obtained with the standard choice $\mu_h^2>0$. For $W^+ W^-$ production with $\pTveto=25$\,GeV, the increase in the central value of the NNLL+NLO cross section is about 4.8\% (which is of the order as the recently calculated NNLO corrections \cite{Gehrmann:2014fva}).\footnote{There is an ambiguity when choosing $\mu_h^2<0$ related to the fact that the running coupling $\alpha_s(\mu^2)$ has a cut along the negative $\mu^2$ axis. One can either choose the default matching scale above or below the cut, $\mu_h^2=-Q^2\pm i\epsilon$. Our values for the cross section obtained within the MadGraph5\Q{_}aMC@NLO framework correspond to the principal value prescription, while the authors of \cite{Jaiswal:2014yba} adopt the default choice $\mu_h^2=-Q^2-i\epsilon$. At NNLL order, the latter choice yields a result that is 2\% higher (at $\pTveto=25\,{\rm GeV}$) than that obtained with the principal-value prescription. This difference would be reduced at higher orders. A detailed numerical comparison with \cite{Jaiswal:2014yba} further revealed that there was a problem in their implementation of the beam functions. Correcting this, our results are in agreement.} 
We stress, however, that in the case of multi-particle final states such as $W^+W^-$ the hard function depends on several kinematic scales ($\hat s$ and $\hat t$ in the present case), some of which are time-like and some of which are space-like. Unfortunately, it is impossible to adopt a suitable choice of the hard matching scale, which would map the hard function onto a Euclidean quantity, such that all $(\alpha_s\pi^2)^n$ terms can be resummed by means of RG evolution equations. It is therefore not clear whether the convergence of the perturbation series can be improved by using the choice $\mu_h^2<0$. This problem was discussed in detail in the context of Higgs plus jet production in \cite{Becher:2014tsa}. Even though the convergence in the right panels of Figure~\ref{WWZRes} looks somewhat better than in the plots shown on the left, we have decided to adopt the conventional prescription $\mu_h>0$ for the hard matching scale. Perhaps a more conservative way to assess the scale uncertainty would be to allow for arbitrary complex scale choices $Q/2<|\mu_h|<2Q$ and then give the resulting uncertainty, as was recently proposed in \cite{Jaiswal:2014cna}. 

For Higgs production, the resummation of jet-veto logarithms was performed to higher accuracy by including the two-loop hard and beam functions as well as the RG evolution factor at approximate N$^3$LL order \cite{Becher:2013xia}. The only missing ingredients for full N$^3$LL+NNLO accuracy are the three-loop anomaly exponent and the four-loop cusp anomalous dimension, whose effects have been estimated and included in the error budget. It was observed in this reference that the two-loop beam functions decrease the cross section, and we expect a similar effect in the present case. In the future, it should be possible to reach the same level of accuracy also for $W^+ W^-$ production and related processes. The corresponding two-loop hard functions can be extracted from the two-loop virtual corrections, which have recently been obtained in \cite{Gehrmann:2014fva,Caola:2014iua}. The product of beam functions integrated over rapidity could be extracted numerically from NNLO fixed-order codes for $Z$-boson production such as \cite{Catani:2009sm,Li:2012wna}, following the procedure employed in \cite{Becher:2013xia}. This is sufficient to obtain the inclusive $W^+ W^-$ cross section, while a two-loop computation of the beam functions would be required for more exclusive cross-section predictions. Once (approximate) N$^3$LL+NNLO predictions for the $W^+ W^-$ cross sections are available, the above-mentioned ambiguities related to the choice of the hard matching scale will be reduced significantly.

\subsection{Fixed-Order Results and Matching\label{sec:matching}}

\begin{figure}[t!] 
\begin{center}      
   \mbox{\includegraphics[width=0.4\textwidth]{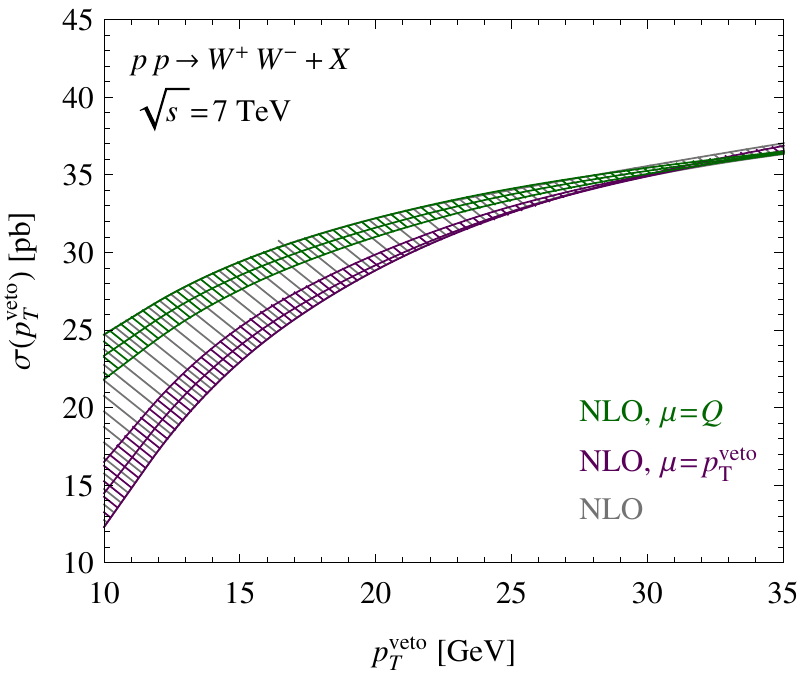}}
   \hspace{20px} 
   \mbox{\includegraphics[width=0.4\textwidth]{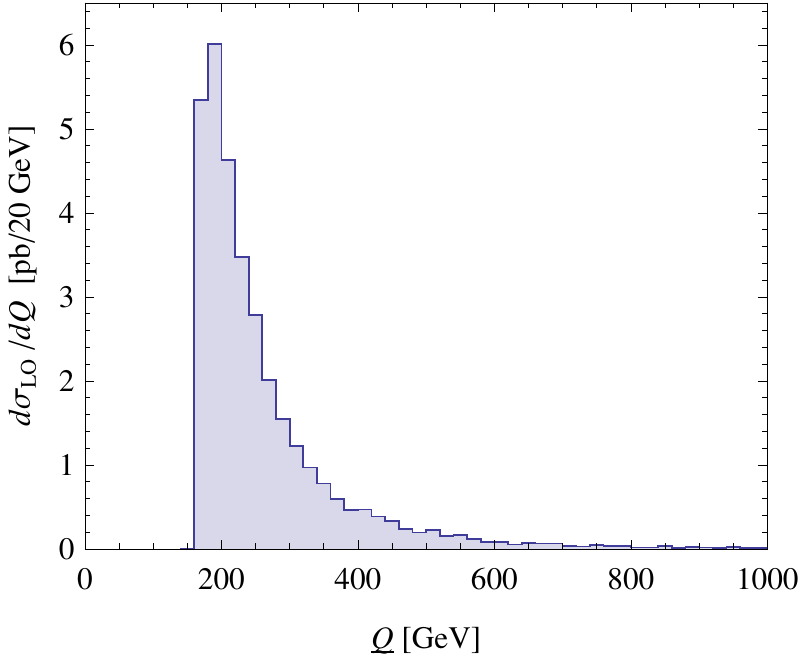}}   
\caption{Left: NLO predictions for the $W^+ W^-$ production cross section obtained with a conservative estimate of scale uncertainties (grey), and with scale variations about high (green) and low (magenta) default values; see text for further information. Right: Kinematic distribution in the variable $Q$ of the leading-order cross section.}
\label{WWNLO}
\end{center}
\end{figure}

In order to obtained the best possible predictions, we need to match our resummed results for the cross sections with fixed-order expressions at NLO. The scale dependence of the NLO expression for the for the $W^+ W^-$ production cross section at $\sqrt{s}=7$\,TeV is shown in the left panel in Figure~\ref{WWNLO}. We set the factorization and renormalization scales equal ($\mu=\mu_r=\mu_f$) and vary them from $\mu=\pTveto/2$ up to $\mu=2Q$. This is a much larger scale variation than is usually considered, but this wider range seems appropriate since the problem at hand involves physics at both scales. For comparison, we also show the bands one would obtain from a variation of $\mu$ by a factor of 2 around either a high default value $\mu=Q$ or a low default value $\mu=\pTveto$. Our broad scale variation is obviously more conservative, since it covers both options. 
Nevertheless, fixed-order computations usually adopt the high scale $\mu=Q$ as the default value, and from Figure~\ref{WWZRes} it appears that such a choice indeed leads to smaller higher-order corrections. A similar behavior is found for all cases studied in this paper. The invariant-mass distribution of the $W$-boson pair is shown in the right panel of Figure~\ref{WWNLO}. Defining the average hard scale $\tilde Q$ by the median value of this distribution, one obtains $\tilde Q=222\,{\rm GeV}$. This value will be useful in our phenomenological discussion below.

As discussed earlier and shown in (\ref{eq:match}), in Scheme A this matching is purely additive, i.e.\ 
\begin{equation}
   \sigma_{\text{NNLL+NLO}} = \sigma_{\text{NNLL}}(\mu,\mu_h) 
    + \Big( \sigma_{\text{NLO}}(\mu_m)
    - \sigma_{\text{NNLL}}(\mu_m)\big|_{\text{expanded to NLO}} \Big) \,.
\end{equation}
The expansion of the resummed result is obtained by performing the reweighting with the reweighting factor expanded to NLO. If the resummation is performed with NNLL accuracy (or higher), the matching correction inside the parentheses is power suppressed in $\pTveto/Q$. Note that we are free to use a different scale $\mu_m$ for the matching correction than for the resummed result, since the power corrections in $\pTveto/Q$ must be separately scale invariant. To obtain our uncertainty bands, the scales $\mu$, $\mu_h$ and the matching scale $\mu_m$ are all varied independently. We then add the resulting uncertainties quadratically. We choose the number of flavors for the resummed results as $n_f=5$, but since MadGraph5\Q{_}aMC@NLO cannot produce five-flavor NLO results for $W^+W^-$ due to the presence of top-quark resonant contributions in the NLO corrections, we calculate the matching corrections with $n_f=4$ light flavors.

\begin{figure}[t!] 
\begin{tabular}{rr}
   \mbox{\hspace{8mm}\includegraphics[width=0.4\textwidth]{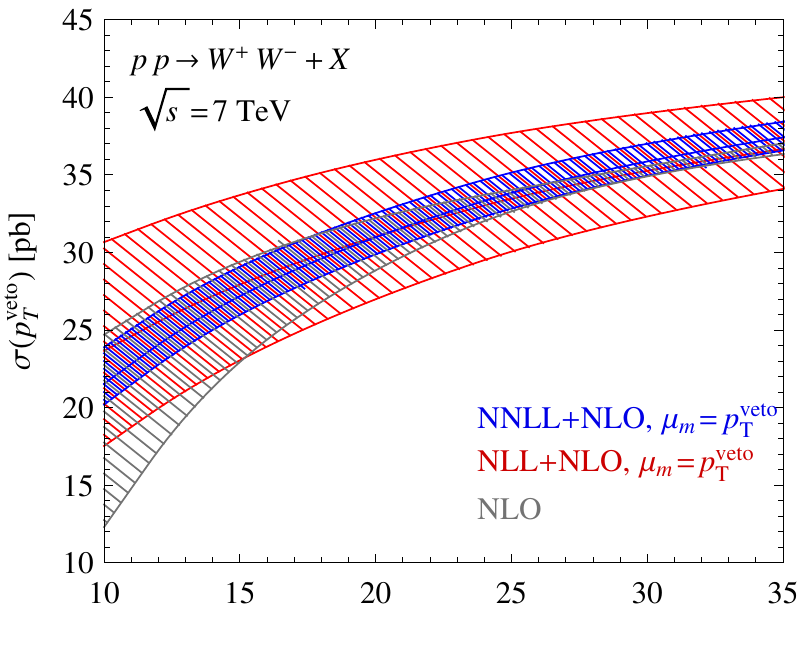}}&   
   \mbox{\includegraphics[width=0.4\textwidth]{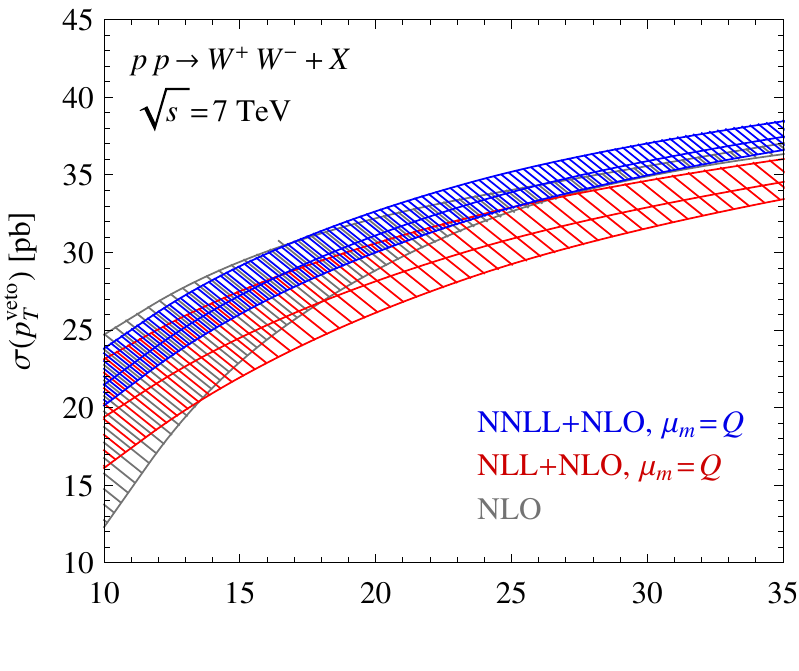}}\\[-6mm]
   \mbox{\includegraphics[width=0.435\textwidth]{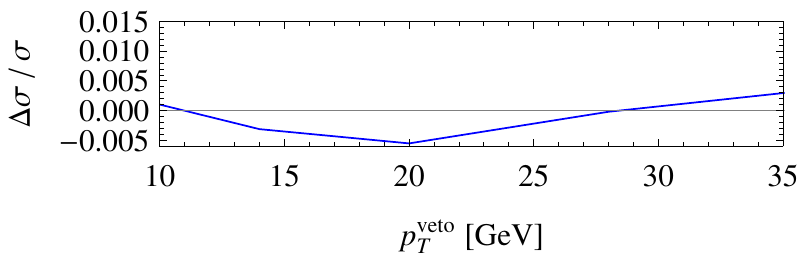}}& 
   \mbox{\includegraphics[width=0.435\textwidth]{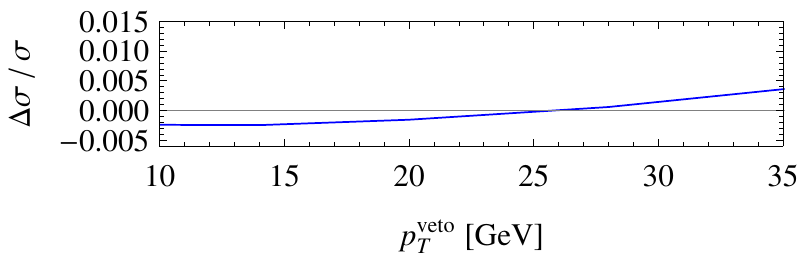}}\\
\end{tabular}
\caption{Resummed and matched predictions for the $W^+ W^-$ production cross section (obtained by varying the matching scale about the default value $\mu_m=\pTveto$ and $\mu_m=Q$) compared with the fixed-order result at NLO. The panels below the plots indicate the relative size of the power-suppressed matching corrections at NNLL order.}
\label{WWres}
\end{figure}

While the appropriate scale choice is clear for the case of the beam functions which describe emissions near the scale $\pTveto$, the correct choice of $\mu_m$ is not immediately obvious, because the matching corrections receive contributions associated with both the low and the high scale. The result for the cross section obtained with a high and a low matching scale is shown in Figure~\ref{WWres}, along with the corresponding relative size of the NNLL matching corrections. The matching corrections are well-behaved in both cases. They are very small at the low $\pTveto$ values shown in Figure~\ref{WWres} and are therefore difficult to extract numerically. At larger values of $\pTveto$ they grow linearly up to 3\% at $\pTveto=80$\,GeV. At NNLL order, the matching corrections are small enough that they could be safely ignored for values up to $\pTveto=35$\,GeV. At NLL order, on the other hand, not all leading-power NLO contributions are included in the resummed result, and therefore the predictions depend strongly on the matching scale $\mu_m$. Figure~\ref{WWZRes} shows that the NNLL results lie rather close to the NLO results at the high scale $\mu=Q$. Since, as we have pointed out above, the fixed-order perturbative expansion appears to work better with a high scale choice, we adopt $\mu_m=Q$ as our default matching scale for all later predictions. 

\begin{figure}[t!] 
\centering
\begin{tabular}{rr}      
   \mbox{\includegraphics[width=0.4\textwidth]{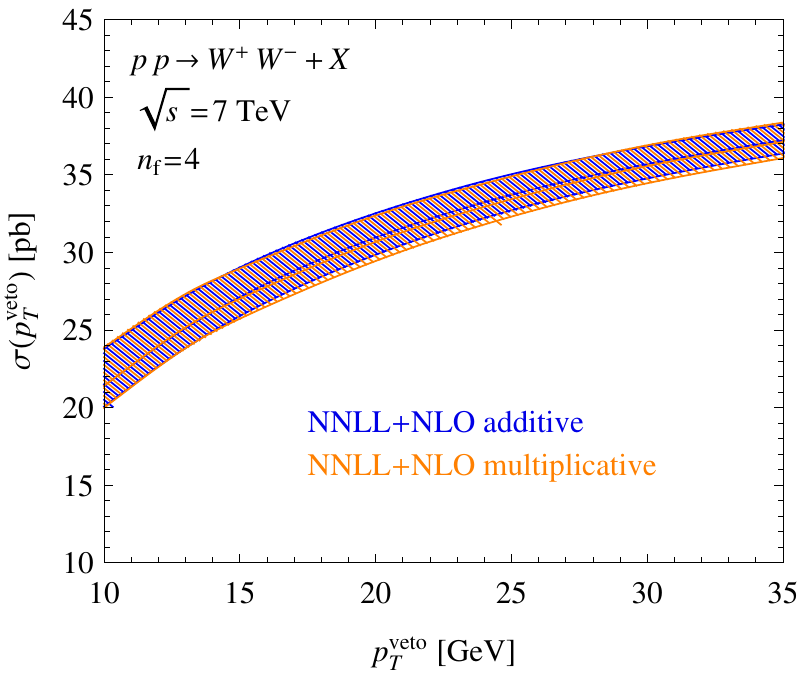}} ~&~
   \mbox{\includegraphics[width=0.4\textwidth]{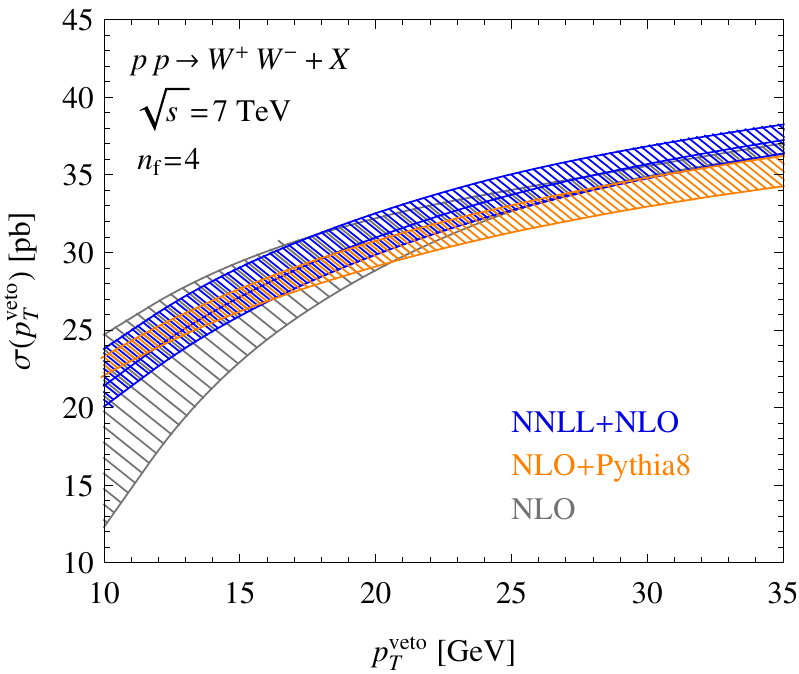}}
\end{tabular}
\caption{Left: Comparison of the resummed and matched NNLL+NLO predictions for the $W^+ W^-$ cross section obtained in Scheme A (additive matching) with Scheme B (multiplicative matching). Right: Comparison of the NNLL+NLO predictions with the NLO result matched to Pythia using aMC@NLO.}
\label{MatchShower}
\end{figure}

In Scheme B, we do not have the freedom to choose the matching scale separately, since the matching corrections are not separated out, see (\ref{eq:reweight}). Numerically, we find that the results of Scheme A and Scheme B are almost indistinguishable, as can be seen in the left panel of Figure~\ref{MatchShower}. In the right panel of the same figure we show a comparison between our NNLL+NLO prediction for the $W^+ W^-$ cross section and the result obtained after combining the NLO prediction with a parton shower using the MC@NLO prescription \cite{Frixione:2002ik}. We observe that the latter prediction is lower than our result, in particular at higher values pf $\pTveto$. This is astonishing at first sight, since one would expect that showering does not affect the cross section at higher $\pTveto$ values. However, because the shower is unitary any change of the cross section at low transverse momenta must be accompanied by a compensating change at higher transverse momenta. Looking at the cross section as a function of the $p_T$ of the leading jet, we find that the showered NLO result is higher than pure NLO result for all $p_T^j>20\,{\rm GeV}$, so that the integral of the cross section for $p_T>20\,{\rm GeV}$ is larger than the fixed-order result. After unitarization, this in turn implies that the jet-veto cross section, which is the integral $0\leq p_T\leq 20\,{\rm GeV}$, is lower than the fixed-order result. The use of a matched parton shower therefore underestimates the jet-veto cross section. In contrast, we find that our NNLL+NLO resummed prediction lies closer to the fixed-order result indicated by the grey band. Genuine resummation effects are small as long as the fixed-order result for the cross section is computed with a high value $\mu\sim Q$ of the renormalization scale.

\subsection{Multiple Bosons and Cross Section Ratios}

\begin{figure}[t!] 
\centering
\hspace*{-0.4cm}\begin{tabular}{rrr}      
   \mbox{\includegraphics[width=0.334\textwidth]{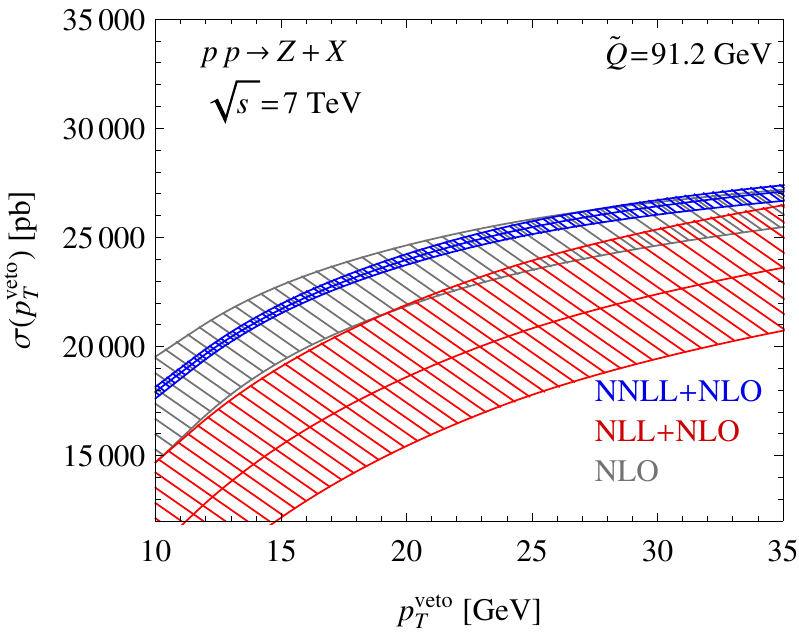}}&
   \mbox{\includegraphics[width=0.310\textwidth]{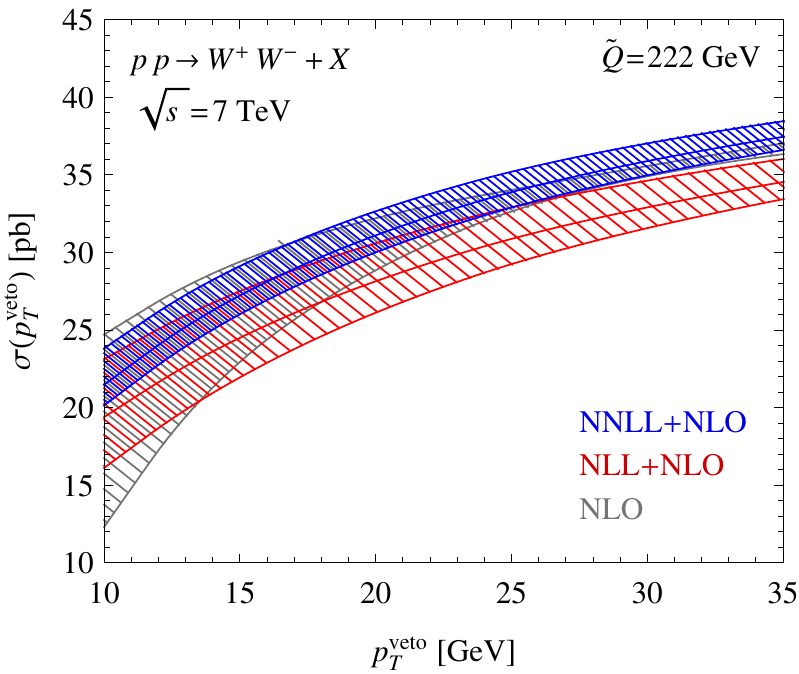}}&
   \mbox{\includegraphics[width=0.319\textwidth]{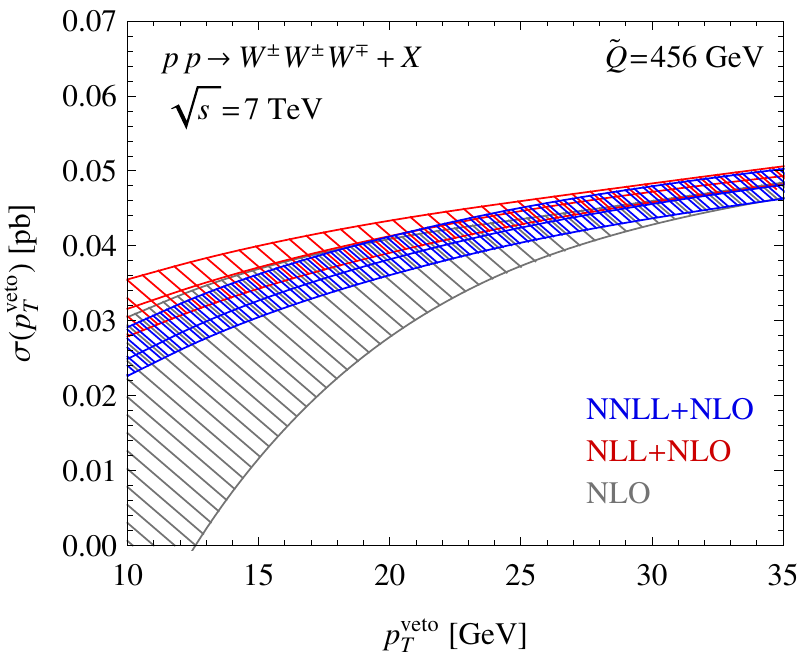}}\\[-6mm]
   \mbox{\includegraphics[width=0.315\textwidth]{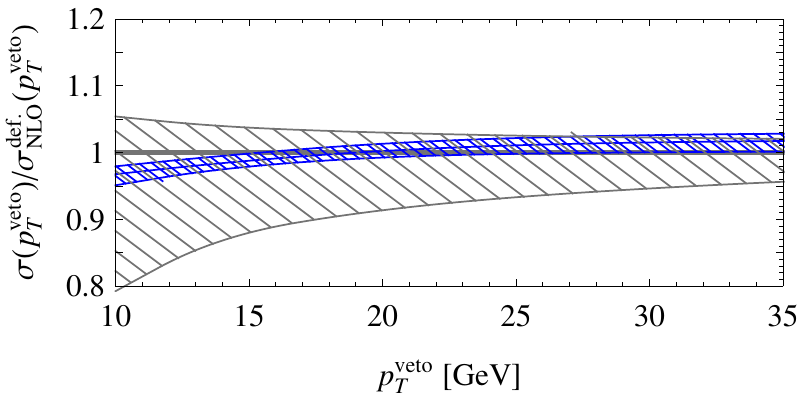}}&  
   \mbox{\includegraphics[width=0.315\textwidth]{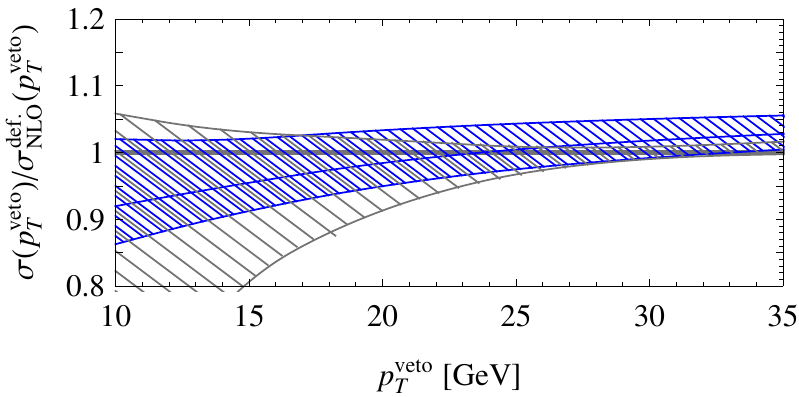}}&
   \mbox{\includegraphics[width=0.315\textwidth]{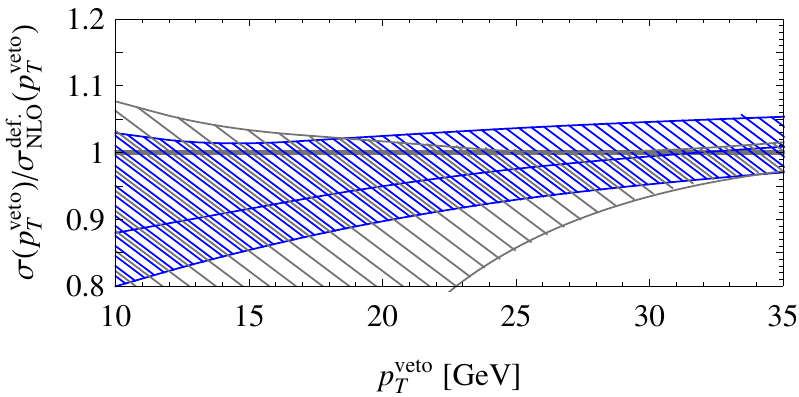}}\\
\end{tabular}
\vspace{-3mm}
\caption{Resummed and matched predictions for the cross sections for $Z$, $W^+ W^-$, and $W^+ W^- W^\pm$ production, compared with NLO fixed-order predictions. The lower panels show the ratio of the cross section to the default NLO value with scale choice $\mu=Q$.}
\label{multiple}
\end{figure}

We are now ready to present our final results for a couple of interesting production cross sections involving multiple electroweak gauge bosons. In Figure~\ref{multiple}, we show predictions for the $Z$, $W^+ W^-$ and $W^+ W^- W^\pm$ production cross sections at the LHC with $\sqrt{s}=7$\,TeV; it would be straightforward to rerun our code at different values of the center-of-mass energy. In each case, we present our resummed and matched predictions at NLL+NLO and NNLL+NLO accuracy and compare them with the fixed-order NLO prediction. Notice that the value of the cross section drops by about a factor $10^3$ with each additional boson. The triple-boson production cross section is tiny, but it constitutes a background to Higgs production in association with a $W^\pm$ and subsequent decay $H\to W^+ W^-$. The fact that we can obtain predictions for three-boson final states without any additional effort nicely demonstrates the power of our automated resummation scheme. 

We find that the scale uncertainties of our NNLL+NLO predictions for $W^+ W^-$ and $W^+ W^- W^\pm$ production are estimated to be of similar size, while we obtain a much smaller uncertainty for the case of $Z$-boson production. This small scale variation should perhaps be taken with a grain of salt. At larger $\pTveto$ values, our resummed cross section becomes similar to the fixed-order result, and its scale variation is similar to the scale variation of the fixed-order cross section obtained by performing a correlated scale variation with $\mu_r=\mu_f$. An independent variation of $\mu_r$ and $\mu_f$, which is standard practice in fixed-order computations, would give an uncertainty that is twice as large. On the other hand, we have checked that the known NNLO corrections for $Z$-boson production are indeed compatible with our small uncertainty band. It is also interesting to note that for $W^+ W^-$ production the scale uncertainties of the fixed-order prediction obtained from correlated and independent variations of $\mu_r$ and $\mu_f$ are found to be of similar size. 

We also observe that the scale uncertainties of the fixed-order NLO predictions at small $\pTveto$ values strongly increase with the number of produced bosons. This is not surprising if we consider the relevant scale ratio $\tilde Q/\pTveto$, which governs the size of Sudakov logarithms. Using the median value $\tilde Q$ of the invariant-mass distribution to estimate the hard scale, we find $\tilde Q=M_Z$ for $Z$ production, $\tilde Q\approx 2.8\,M_W$ for $W^+ W^-$ production, and $\tilde Q=5.7\,M_W$ for $W^+ W^- W^\pm$ production. In all cases, the three-momenta at which the bosons are produced scale with the boson mass, but the average scale increases with the number of the produced bosons. Note that after the resummation of Sudakov logarithms has been performed, the width of the uncertainty bands is only weakly dependent on the veto scale. 

\begin{figure}[t!] 
\centering
\begin{tabular}{ccc}    
   \includegraphics[width=0.325\textwidth]{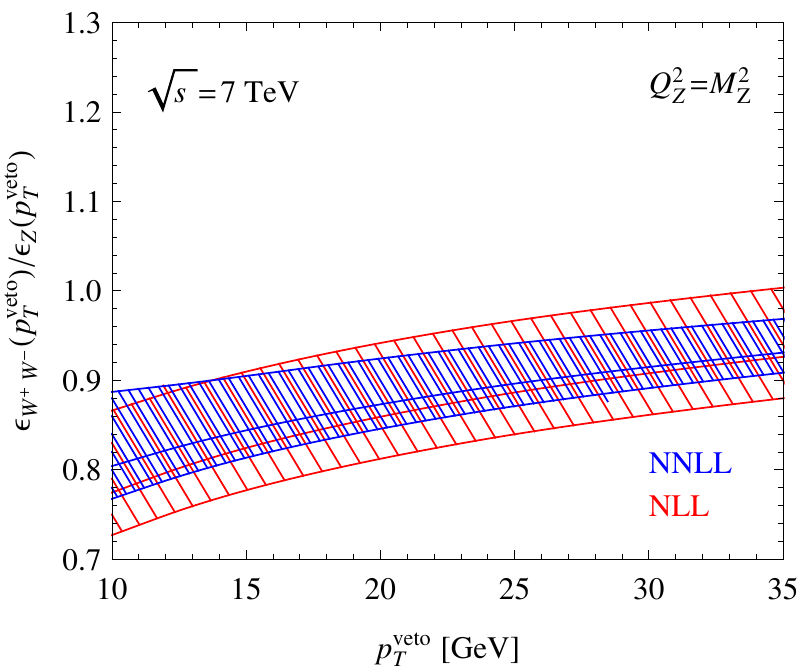} \hspace{-5mm} & 
   \includegraphics[width=0.325\textwidth]{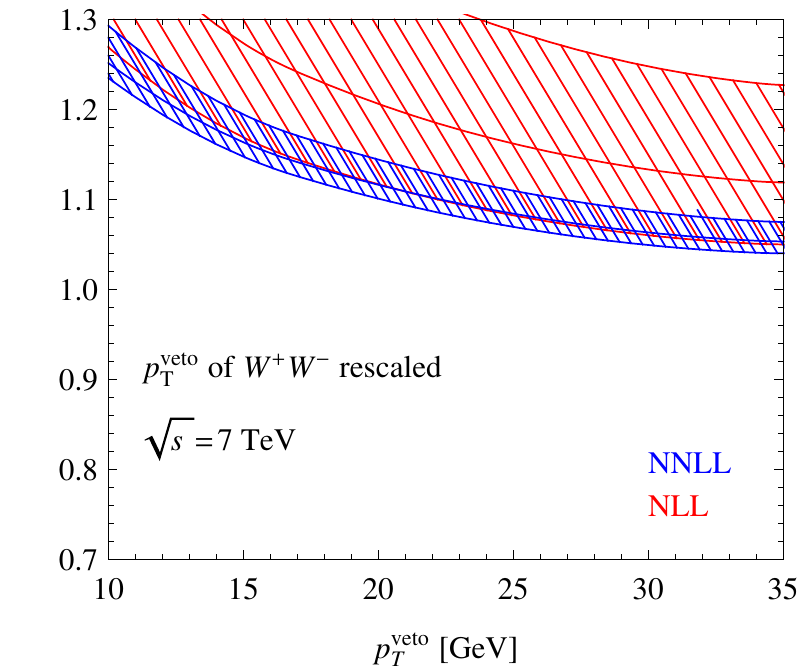} \hspace{-5mm} & 
   \includegraphics[width=0.325\textwidth]{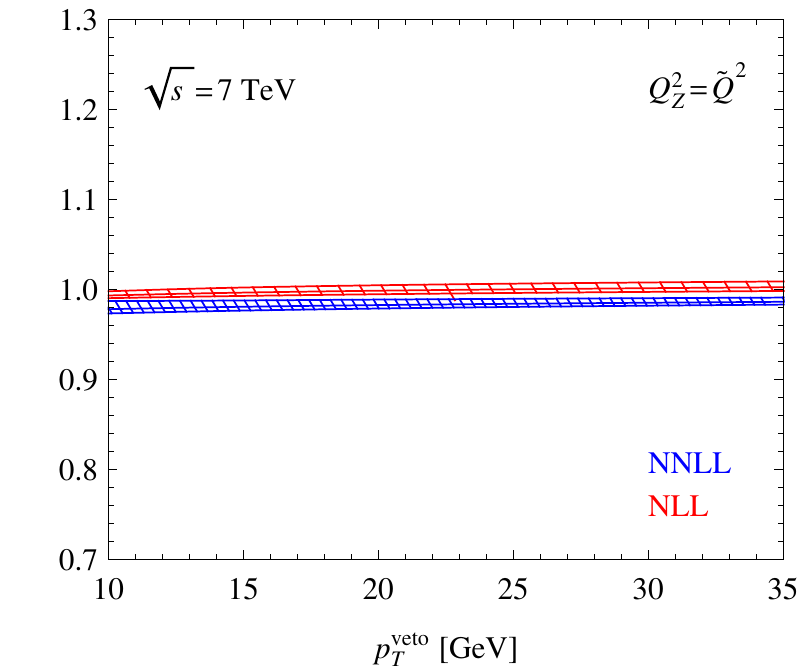}
\end{tabular}
\caption{Resummed predictions for the ratio of the jet-veto efficiencies for $W^+ W^-$ and $Z$-boson production (left). In the middle plot the $\pTveto$ value of the $W^+ W^-$ process is rescaled by a factor $M_Z/(2M_W)$, as proposed in \cite{Monni:2014zra}. The right plot shows the same ratio for $W^+ W^-$ and $Z^*$-boson production, where the off-shell boson has invariant mass $\tilde{Q}_{WW}=222\,{\rm GeV}$. The bands are obtained by varying the low scale $\mu$ about its default value $\mu=\pTveto$, while keeping the hard matching scale $\mu_h$ fixed.}
\label{WWZ}
\end{figure}

The relative perturbative uncertainty of our NNLL+NLO prediction for the $W^+ W^-$ production cross section at $\pTveto=25$\,GeV is $^{+3.9\%}_{-3.0\%}$. It was advocated in \cite{Campbell:2009kg} that taking the ratio of the $W^+ W^-$ and $Z$-boson production cross sections might be a good way to reduce the uncertainty in the prediction of the jet-veto cross sections. This proposal was adopted in the experimental analysis reported in \cite{ATLAS:2012mec}. We have thus studied this cross-section ratio in some detail. We find that the relative uncertainty in the cross-section ratio is $^{+5.2\%}_{-2.8\%}$, which is even slightly larger than the uncertainty in the $W^+ W^-$ production cross section itself. This makes it clear that taking the cross-section ratio does not help reducing the perturbative uncertainties, the reason being that the scale uncertainties are much smaller for $Z$-boson production than for $W^+ W^-$ production. Even though the beam functions are the same in both cases, the cross sections involve different hard functions and RG evolution factors, which spoils the cancellation. We will now explain how an improved relation between the two production channels can be obtained, which only suffers from very small theoretical uncertainties. In a first step, it is useful to consider the jet-veto efficiencies defined as $\sigma(\pTveto)/\sigma$ instead of the cross sections $\sigma(\pTveto)$ themselves, because then the virtual corrections encoded in the hard functions largely drop out (even though this cancellation cannot be exact, since the hard corrections do not factor out of the total cross section). The inclusive cross section $\sigma$ is evaluated at the hard scale $\mu_h$. We use the NLO (LO) cross section together with the NNLL (NLL) approximation of $\sigma(\pTveto)$. Our resummed predictions for the ratio of the jet-veto efficiencies for $W^+ W^-$ and $Z$ production are shown in the left plot in Figure~\ref{WWZ}. By construction, the relative uncertainties from varying $\mu$ in this ratio are the same as in the ratio of the veto cross sections. In order to obtain more accurate predictions one needs to ensure that the RG evolution factors cancel out in the ratio. This can be accomplished by considering the ratio of the $W^+ W^-$ cross section to the $Z^*$ production cross section with an off-shell $Z^*$ boson with invariant mass squared $q^2=\tilde{Q}^2$, where $\tilde{Q}\approx 222\,{\rm GeV}$ is the median of the invariant-mass distribution for the $W^+ W^-$ final state shown in Figure~\ref{WWNLO}. The corresponding ratio of efficiencies is shown in the right plot in Figure~\ref{WWZ}. It is close to~1 and exhibits very small scale uncertainties. A different way of relating $Z$ and $W^+ W^-$ production cross sections was proposed in \cite{Monni:2014zra}. These authors rescale the $\pTveto$ value used in the $W^+ W^-$ process by a factor $M_Z/(2M_W)$ before relating it to the $Z$-boson production process. This rescaling is chosen such that the Sudakov logarithms have a similar size in the two cases. While \cite{Monni:2014zra} finds a nice agreement for the NLO efficiencies obtained using this rescaling prescription, it is clear that the relation cannot be exact, since QCD is not scale invariant. Furthermore, the agreement becomes worse if one rescales the $\pTveto$ value with the more appropriate factor $M_Z/\tilde{Q}$. In the middle plot of Figure~\ref{WWZ}, we show the corresponding ratio of efficiencies, which suffers from sizable scale uncertainties.

\subsection{Experimental Cuts}

\begin{figure}[t!] 
\begin{center}
\begin{tabular}{r}   
   \mbox{\includegraphics[width=0.45\textwidth]{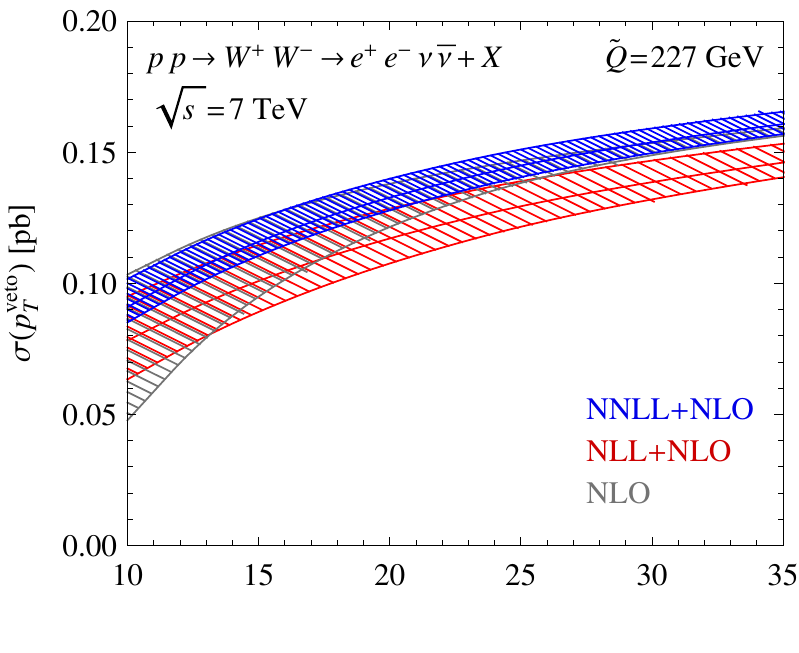}} \\[-6mm]
   \mbox{\includegraphics[width=0.473\textwidth]{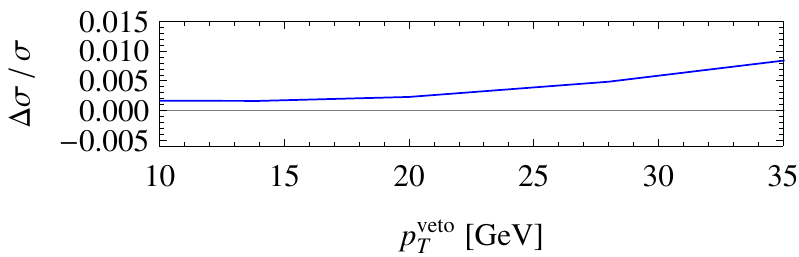}}
\end{tabular}
\caption{Resummed and matched predictions for the $pp\to W^+ W^- +X\to e^+ e^-\nu\bar\nu+X$ cross section with the cuts on the leptonic final state described in the text.}
\label{WWdec}
\end{center}
\end{figure}

An important advantage of our framework is that we can include the decay of electroweak bosons, together with cuts on the leptonic final state. In the experimental measurements of $W^+W^-$ production, candidate events are selected with two opposite-sign charged leptons, electrons or muons, and missing transverse momentum coming from the neutrinos in $pp\to W^+ W^- +X\to l\,\nu\,l'\nu'+X$. To account for the detector geometry and to suppress the background from Drell-Yan and top production, a number of cuts are applied to the final state in addition to the jet veto. For example, the ATLAS analysis \cite{ATLAS:2012mec} imposes the following cuts in the $e^+e^-$ channel:
\begin{enumerate}
 \item lepton $p_T>20\,{\rm GeV}$ 
 \item leading lepton $p_T>25\,{\rm GeV}$ 
 \item lepton pseudorapidity $\eta_e<1.37$ or $1.52<\eta_e<2.47$
 \item dilepton invariant mass $m_{e^+e^-}>15\,{\rm GeV}$ and $|m_{e^+e^-}-m_Z|>15\,{\rm GeV}$\label{dileptonmass}
\end{enumerate} 
The cuts applied in the $\mu^+\mu^-$ channel are fairly similar, while those on the mixed final states $e^\pm\mu^\mp$ are looser, because they have much smaller Drell-Yan background. In Figure~\ref{WWdec}, we show the cross section for the production and decay $pp\to W^+ W^- +X\to e^+ e^-\nu \bar\nu+X$ in the presence of these cuts as a function of the jet-veto scale. The experimental analysis in \cite{ATLAS:2012mec} uses the anti-$k_T$ algorithm with $R=0.4$ and fixed $p_T^\text{veto}=25\,{\rm GeV}$. Comparing this figure with the lower plots in Figure~\ref{WWZRes}, we see that the uncertainties of the cross section are similar to the inclusive case and that the matching corrections remain small also in the presence of the cuts.

The experimental analysis \cite{ATLAS:2012mec} imposes a few additional cuts, in particular a minimum total transverse momentum of the two charged leptons $p_T^{e^+e^-}\,>30 \, {\rm GeV }$ and minimum requirements on the missing transverse momentum $p_{T, \text{Rel}}^{\nu\bar\nu}\,>45\, {\rm  GeV}$.\footnote{The exact definition of $p_{T, \text{Rel}}^{\nu\bar\nu}$ is more involved, see\cite{ATLAS:2012mec}.} 
The cut $p_T^{e^+e^-}\,>30 \, {\rm GeV }$ is somewhat problematic for the theoretical analysis, especially when it is applied to predict the $Z$-boson background to $W^+ W^-$ production. The difficulty is that we must make sure that the leptonic cuts do not (strongly) affect the hadronic final state. In the case of $Z$ production the $p_T^{e^+e^-}$ is equal (and opposite) to the transverse momentum $p_T^X$ of the hadronic final state. Imposing a lower bound on $p_T^{e^+e^-}$ is the same as imposing a lower bound on $p_T^X$. This interferes with the jet-veto cut which at NLO corresponds to an upper cut on $p_T^X$. The factorization formula in \cite{Becher:2012qa} does allow for additional cuts on $p_T^X$ in the presence of the jet veto, but the relevant beam functions would be more complicated than those needed without such cuts. For the $W^+ W^-$ production process, the quantity 
$p_T^{e^+e^-}$ is not directly related to $p_T^X$ because of the presence of the neutrinos, but the corresponding cut still affects the low-$p_T^X$ region.

\subsection{Difficulties Associated with Photons}

Our framework cannot immediately be applied to processes involving photons. The reason is that photons are massless particles and have hadronic substructure. At high energies, a photon thus needs to be treated as a photon jet, or more precisely a photon surrounded by some hadronic radiation. In fact, many photon-isolation requirements necessitate fragmentation functions. This can be avoided using the photon isolation proposed by Frixione \cite{Frixione:1998jh}, but also in this case the photon has a partonic content and a proper description needs to take into account partons emitted collinear to the photon. This implies that our factorization theorem does not apply, since it assumes that all energetic radiation is collinear to the beam. The photon isolation introduces new small scales to the problem (e.g.\ the hadronic energy around the photon), which give rise to additional large logarithms not associated with the jet veto.

\begin{figure}[t!] 
\begin{center}     
   \mbox{\includegraphics[width=0.4\textwidth]{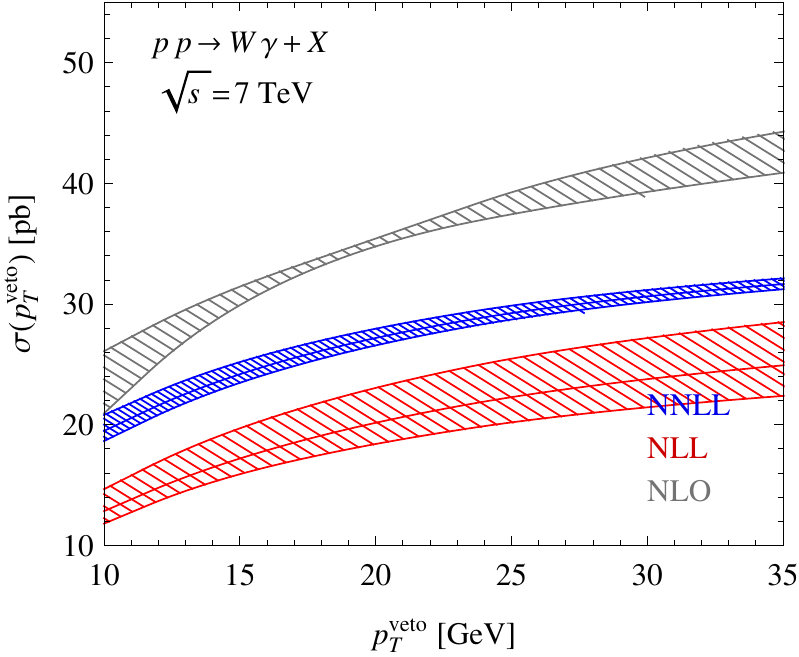}} 
   \hspace{20px} 
   \mbox{\includegraphics[width=0.4\textwidth]{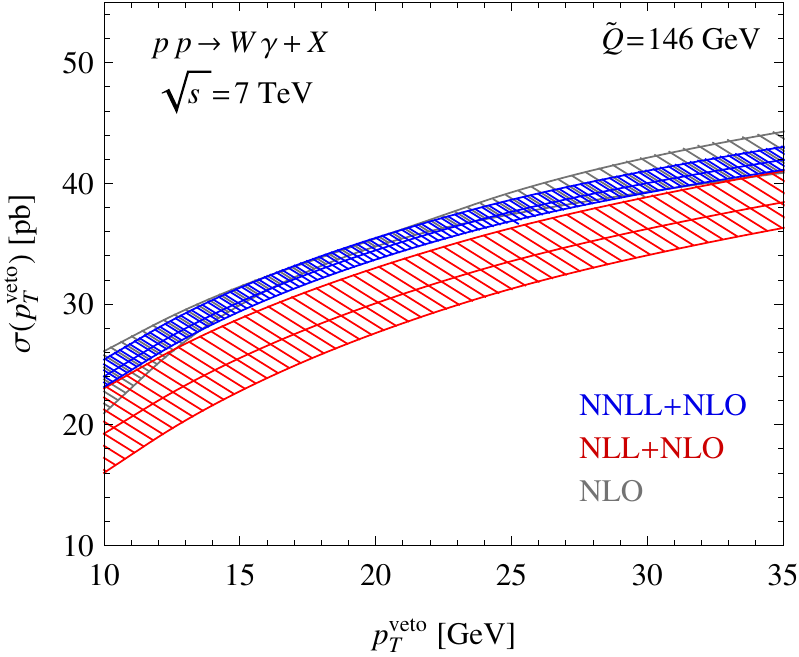}} 
\caption{Theoretical predictions for $W\gamma$ production obtained from our resummation scheme. The left plot shows the resummed results without matching to NLO, while the right plot shows the results obtained after the matching has been performed. A proper treatment of production processes with high-energy photons in the final state would require a generalization of the factorization formula (\ref{eq:sigma}).}
\label{Wg}
\end{center}
\end{figure}

It is nevertheless interesting to see what happens when we apply our resummation scheme to a process involving photons. To this end, we consider $W^\pm\gamma$ production using the same setup as before ($\sqrt{s}=7$\,TeV, $R=0.4$, $n_f=4$) and imposing the isolation requirement proposed in \cite{Frixione:1998jh}, with associated parameters $R_0^\gamma =0.4$, $x_n=1.0$ and $\epsilon_\gamma=1.0$. The corresponding results are shown in Figure~\ref{Wg}. The $pp\to W\gamma$ process suffers from very large NLO corrections (the LO results are similar to the NLL result). The resummed results, on the other hand, are not very different from the LO predictions, so that the matching corrections are huge, indicating that there are indeed other sources of large corrections in this process. Likely these arise due to Sudakov effects associated with photon isolation. However, even the logarithms associated with the jet veto have a more complicated structure once a process involves partons collinear to the photon directions, which becomes possible at NLO. It would be interesting to analyze such photon processes in the context of SCET. In its present implementation our method does not resum all large corrections in these cases.

\section{Conclusion}
\label{sec:conclusion}

Higher-order logarithmic resummations in collider physics, both in SCET and using traditional methods, are typically done on a case-by-case basis, similar to the way fixed-order calculations were performed a few years ago. In the meantime, several groups have automated NLO computations in a variety of computer codes. This automation saves time, reduces the possibilities for mistakes and offers the flexibility to also study effects beyond the Standard Model. It is desirable to have the same level of automation for higher-order resummations of large logarithmic corrections. In the present paper, we have achieved this goal for electroweak-boson production cross sections in the presence of a jet veto, at NNLL+NLO accuracy. This combination is natural because in the Sudakov region, where $\ln(Q/\pTveto)\sim 1/\alpha_s$, NNLL logarithmic terms have the same parametric scaling as NLO corrections in a region where there are no large logarithms. In contrast, taking resummation effects into account using a parton shower gives a lower parametric accuracy, and the unitarization inherent in the shower approach can sometimes be problematic. In the case of the jet-veto cross section for $W^+ W^-$ production, for example, unitarization leads to cross sections that are systematically lower than the NNLL+NLO results. 

Resummations are relevant in kinematical configurations which are close to the Born-level kinematics and can therefore be obtained by reweighting Born-level cross sections with appropriate factors. The most complicated ingredient for NNLL resummations are the one-loop hard functions, which encode the virtual corrections. Their computation has been automated, and we use the MadGraph5\Q{_}aMC@NLO framework to obtain the hard function required for our analysis. We have also presented a modified scheme, in which the beam functions accounting for collinear emissions and the matching onto fixed-order results is automated and performed using existing fixed-order codes. This is possible, because the hard function and the resummation of large logarithms are just overall factors in the differential cross section. 

We have used our method to perform a detailed analysis of resummation effects for the $W^+ W^-$ pair production cross section, for which experimental measurements found a slight excess compared to theoretical predictions based on NLO computations matched to parton showers. We observe that the NLO result with a high value of the renormalization and factorization scales $\mu_r \sim \mu_f \sim Q$ is in good agreement with the NNLL+NLO resummed predictions, while the results obtained with a matched parton shower are systematically lower. This effect, together with the positive NNLO corrections to the total rate which are now known, helps to bring the Standard Model prediction into better agreement with the measurements. It would be important to include the two-loop virtual corrections into the resummation and to also compute and include two-loop beam functions. This improvement, which is beyond the scope of the present work, would lead to very precise predictions, which could be directly compared with the experimental results. This level of accuracy has already been achieved in Higgs production by extracting the beam functions numerically. It was found that the two-loop corrections to the beam functions were sizable, because they are enhanced by logarithms of the jet radius. In Higgs production, the NNLO corrections to the hard function increase the cross section, while the two-loop beam functions lower it. We expect the same behavior for the $W^+ W^-$ case, and it will be interesting to see the combined effect of these improvements on the final predictions. Also, at NNLO the $gg$ channel starts to contribute to $W^+ W^-$ production and could give rise to important corrections. Since this channel has already been implemented into the MadGraph5\Q{_}aMC@NLO framework, it will be straightforward to perform the corresponding resummation using our method.

It would also be interesting to generalize our methods to processes with jets in the final state. In addition to hard, beam, and soft functions, these processes involve jet functions describing the energetic final-state radiation. Furthermore, the hard function then has non-trivial color structure. Existing programs which compute virtual corrections for NLO processes currently only supply squared matrix elements summed over colors, but they can be modified to provide the color information needed for SCET-based resummation. This color structure is then contracted with the color structure of the soft function after RG evolution. The soft, beam and jet functions will in general need a separate calculation. However, since the jet and beam functions are two-point functions and the soft function is given by a single emission from eikonal lines, these computations are much simpler than full-fledged real-emission computations and could be automated as well. We are confident that such automated resummations will become available in the future and provide higher-order logarithmic resummations for a much wider range of observables. 

\vspace{2mm}
{\em Acknowledgments:\/}
We thank Prerit Jaiswal and Takemichi Okui for discussions and for their help in performing a detailed numerical comparison to their results. T.B.\ and L.R.\ are supported by the Swiss National Science Foundation (SNF) under grant 200020\Q{_}153294 and acknowledge support from the Munich Institute for Astro- and Particle Physics (MIAPP) and from the Mainz Institute for Theoretical Physics (MITP). The research of M.N.\ is supported by the Advanced Grant EFT4LHC of the European Research Council, the Cluster of Excellence {\em Precision Physics, Fundamental Interactions and Structure of Matter\/} (PRISMA -- EXC 1098), and grant 05H12UME of the German Federal Ministry for Education and Research. 

\begin{appendix}
\numberwithin{equation}{section}
\section{Ingredients Required for NNLL Resummation}

In the following we list the expressions used in (\ref{pdfmatch}) and (\ref{eq:Prefactor}). The one-loop kernel functions $\bar I_{j\leftarrow i}$ for jet-veto cross sections have the same form as those relevant for transverse-momentum resummation. They were first obtained in \cite{Becher:2011xn,Becher:2012qa} and read
\begin{equation}\label{barIexp}
   \bar I_{j\leftarrow i}(z,\pTveto,\alpha_s) 
   = \delta(1-z)\,\delta_{ji} -\frac{\alpha_s}{4\pi} \left[ 
    {\cal P}_{j\leftarrow i}^{(1)}(z)\,\frac{L_\perp}{2} - {\cal R}_{j\leftarrow i}(z) \right] 
    + {\cal O}(\alpha_s^2) \,,
\end{equation}
where we have defined the abbreviation $L_\perp=2\ln(\mu/\pTveto)$. The one-loop DGLAP splitting functions read
\begin{equation}\label{APkernels}
\begin{aligned}
   {\cal P}_{q\leftarrow q}^{(1)}(z) 
   &= 4C_F \left( \frac{1+z^2}{1-z} \right)_+ , \qquad
   {\cal P}_{q\leftarrow g}^{(1)}(z) 
   = 4T_F \left[ z^2 + (1-z)^2 \right]\,,\\
   {\cal P}_{g\leftarrow g}^{(1)}(z) 
   &= 8C_A \left[ \frac{z}{\left(1-z\right)_+} + \frac{1-z}{z} + z(1-z) \right]
    + 2\beta_0\,\delta(1-z) \,, \\
   {\cal P}_{g\leftarrow q}^{(1)}(z) 
   &= 4C_F\,\frac{1+(1-z)^2}{z}\,,
\end{aligned}
\end{equation}
and the remainder functions are
\begin{equation}
\begin{aligned}
   {\cal R}_{q\leftarrow q}(z) 
   &= C_F \left[ 2(1-z) - \frac{\pi^2}{6}\,\delta(1-z) \right] , \qquad
   {\cal R}_{q\leftarrow g}(z) 
   = 4T_F\,z(1-z)\,,\\
   {\cal R}_{g\leftarrow g}(z) &= - C_A\,\frac{\pi^2}{6}\,\delta(1-z) \,, \qquad
   {\cal R}_{g\leftarrow q}(z) = 2C_F z \,.
\end{aligned}
\end{equation}

To distinguish between the gluon and quark channel in the anomalous dimensions we use the notation $\gamma^i$ with $i= g$ for the gluon and $i= q$ for the quark channel. The relevant quadratic Casimir operators $C_i$ are $C_g=C_A$ and $C_q=C_F$. The RG evolution factor for the hard function in (\ref{eq:Prefactor}) takes the general form 
\begin{equation}\label{CVsol}
   U_i(Q^2,\mu_h,\mu)
   = \exp\left[ 4 C_i\,S(\mu_h,\mu) - 4a_{\gamma^i}(\mu_h,\mu) \right]
    \left( \frac{Q^2}{\mu_h^2} \right)^{-2C_i\,a_\Gamma(\mu_h,\mu)}\,.
\end{equation}
The Sudakov exponent $S$ and the exponents $a_n$ are given by \cite{Neubert:2004dd}
\begin{equation}\label{RGEsols}
\begin{aligned}
   S(\mu_h,\mu) 
   &= \frac{\Gamma_0}{4\beta_0^2}\,\Bigg\{
    \frac{4\pi}{\alpha_s(\mu_h)} \left( 1 - \frac{1}{r} - \ln r \right)
    + \left( \frac{\Gamma_1}{\Gamma_0} - \frac{\beta_1}{\beta_0}
    \right) (1-r+\ln r) + \frac{\beta_1}{2\beta_0} \ln^2 r \\
   &\hspace{1.6cm}\mbox{}+ \frac{\alpha_s(\mu_h)}{4\pi} \Bigg[ 
    \left( \frac{\Gamma_1\beta_1}{\Gamma_0\beta_0} 
    - \frac{\beta_2}{\beta_0} \right) (1-r+r\ln r)
    + \left( \frac{\beta_1^2}{\beta_0^2} 
    - \frac{\beta_2}{\beta_0} \right) (1-r)\ln r \\
   &\hspace{3.7cm}
    \mbox{}- \left( \frac{\beta_1^2}{\beta_0^2} 
    - \frac{\beta_2}{\beta_0}
    - \frac{\Gamma_1\beta_1}{\Gamma_0\beta_0} 
    + \frac{\Gamma_2}{\Gamma_0} \right) \frac{(1-r)^2}{2} \Bigg] + \dots \Bigg\} \,, \\
   a_\Gamma(\mu_h,\mu) 
   &   = \frac{\Gamma_0}{2\beta_0} \left[ \,\ln\frac{\alpha_s(\mu)}{\alpha_s(\mu_h)}
    + \left( \frac{\Gamma_1}{\Gamma_0} - \frac{\beta_1}{\beta_0} 
    \right) \frac{\alpha_s(\mu) - \alpha_s(\mu_h)}{4\pi} + \dots \right] , 
\end{aligned}
\end{equation}
where $r=\alpha_s(\mu)/\alpha_s(\mu_h)$. A similar expression, with the coefficients $\Gamma_j$ replaced by $\gamma^i_j$, holds for the function $a_{\gamma^i}$. The relevant expansion coefficients of the anomalous dimensions and $\beta$-function can be found, e.g., in \cite{Becher:2009qa}.

The anomaly exponent and the factor $h_i$ are given by \cite{Becher:2010tm}
\begin{equation}\label{Fhexpansions}
\begin{aligned}
   F_i(\pTveto,\mu)
   &= \frac{\alpha_s(\mu)}{4\pi}C_i\,\Gamma_0\,L_\perp
    + \left(\frac{\alpha_s(\mu)}{4\pi}\right)^2 \left[C_i\,\Gamma_0\,\beta_0\,\frac{L_\perp^2}{2} 
    + C_i\,\Gamma_1\,L_\perp + d_{2i}^{\rm veto}(R) \right] , \\
   h_i(\pTveto,\mu)
   &= \frac{\alpha_s(\mu)}{4\pi} \left[C_i\,\Gamma_0\,\frac{L_\perp^2}{4} 
    - \gamma_0^i\,L_\perp \right] \,.
\end{aligned}
\end{equation}
The anomaly coefficient $d_2^{\rm veto}(R)$ given in \cite{Becher:2013xia} is of the form
\begin{equation}\label{notCasi}
   d_{2i}^{\rm veto}(R) = C_i \left[ C_A \left( \frac{808}{27} - 28\zeta_3 \right)
    - \frac{224}{27}T_F n_f \right] - 32\,C_i\,f_i(R) \,,
\end{equation}
where the expansion of $f_i(R)$ for small $R$ reads, in numerical form,
\begin{equation}\label{fR}
\begin{aligned}
   f_i(R) &= - \left( 1.0963\,C_A + 0.1768\,T_F n_f \right) \ln R
    + \left(  0.6106\,C_A -  0.0310\,T_F n_f \right) \\
   &\hspace{4.7mm}\mbox{}+ \left( 0.2639\,C_A-0.8225\,C_i + 0.0221\,T_F n_f \right) R^2\\
   &\hspace{4.7mm}\mbox{}+ \left(- 0.0226\,C_A +0.0625\,C_i- 0.0004\,T_F n_f \right) R^4 + \dots \,.
\end{aligned}
\end{equation} 

\end{appendix}

\end{document}